\documentclass[11pt]{article}
\usepackage{color,graphicx,amssymb,amsfonts}%,showlabels}
\usepackage[mathscr]{eucal}
\title{An Action Principle for the Masses of Dirac Particles}
\author{Felix Finster\thanks{Supported in part by the Deutsche Forschungsgemeinschaft.}
and Stefan Hoch}
\date{December 2007}

\newtheorem{Def}{Definition}[section]
\newtheorem{Thm}[Def]{Theorem}
\newtheorem{Prp}[Def]{Proposition}
\newtheorem{Lemma}[Def]{Lemma}

\newtheorem{Corollary}[Def]{Corollary}

\newcommand{\Proof}{{\em{Proof.}}}
\newcommand{\QED}{\ \hfill $\FBox$ \\[1em]}

\newcommand{\spc}{\;\;\;\;\;\;\;\;\;\;}

\newcommand{\C}{\mathbb{C}}
\newcommand{\R}{\mathbb{R}}
\newcommand{\1}{\mbox{\rm 1 \hspace{-1.05 em} 1}}

\newcommand{\N}{\mathbb{N}}

\newcommand{\Pdd}{\mbox{$\partial$ \hspace{-1.23 em} $/$}}
\newcommand{\slsh}{\mbox{ \hspace{-1.15 em} $/$}}
\newcommand{\qslsh}{\mbox{$q$ \hspace{-1.17 em} $/$}}
\newcommand{\Tr}{\mbox{\rm{Tr}\/}}

\newcommand{\beq}{\begin{equation}}
\newcommand{\eeq}{\end{equation}}
\newcommand{\FBox}{\rule{2mm}{2.25mm}}
\newcommand{\OBox}{\raisebox{.6ex}{\fbox{}}\,}
\newcommand{\hF}{\hat{\mathcal{F}}}

\newcommand{\M}{{\mathcal{M}}}
\newcommand{\tM}{\tilde{\mathcal{M}}}
\newcommand{\hM}{\hat{\mathcal{M}}}
\newcommand{\hN}{\hat{\mathcal{N}}}

\newcommand{\m}{{\mathfrak{m}}}

\begin{document}
\maketitle
\begin{abstract}
A variational principle is introduced which minimizes an action
formulated for configurations of vacuum Dirac seas.
The action is analyzed in position and momentum space.
We relate the corresponding Euler-Lagrange equations to
the notion of state stability. Examples of
numerical minimizers are constructed and discussed.
\end{abstract}

\tableofcontents

\newpage
\section{Introduction}
\setcounter{equation}{0}  \label{sec1}
In the standard model of elementary particle physics many structures
and parameters are built in ad-hoc. In particular, the theory makes no
statement on the masses of the leptons and quarks; the mass parameters
need to be put in empirically from experimental data.
This situation is not quite satisfying, and one would hope for a more
fundamental explanation for the ratios of these masses.
As an approach towards a theory which makes predictions for the
mass parameters, we here set up a variational principle
where we minimize an action formulated for configurations of vacuum Dirac seas.

The minimization problem is nonlocal, nonlinear and involves two or four free
parameters. Therefore, a systematic study of the structure of all minimizers goes beyond
the scope of this paper. But we construct and discuss numerical examples of minimizers.
The technical core of the paper is to compute the Fourier transform of
the action and to relate the regularization on the light cone
to a suitable regularization procedure in momentum space.
This transformation to momentum space simplifies
the numerical analysis. Furthermore, it reveals
a connection between the corresponding Euler-Lagrange equations
and the notion of state stability as
introduced in~\cite[\S5.6]{PFP} for vacuum Dirac sea configurations.

Since the action principle for vacuum Dirac seas seems of interest by itself,
this paper is self-contained and easily accessible. Nevertheless, our action
principle can be seen in a broader context as a modification of the variational
principle introduced in~\cite{PFP} for the fermionic projector in discrete space-time
(see also~\cite{F1} or the review articles~\cite{F2, F5}).
More specifically, we modify the action of~\cite{PFP} such as to make it finite in the
Lorentz invariant setting, treating the singularities on the light cone by a suitable
regularization procedure.
Our method is complementary to that in~\cite[Chapter~4]{PFP}: whereas in~\cite{PFP}
the behavior of the singularities on the light cone is considered in the so-called
continuum limit, we here disregard these singularities completely and concentrate instead on the
regular contribution away from the light cone.
Yet another method is used in~\cite{F4}, where the vacuum Dirac seas are regularized,
and it is analyzed how the operator product~$\M P$ behaves as the regularization
is removed.
These different approaches all complement each other and should eventually
be combined to get a full understanding of the structure of the
minimizers of our variational principle. For an overview of the
different methods we refer to the review paper~\cite{F6}.

The paper is organized as follows. In Section~\ref{sec2} we set up our
variational principle in position space and discuss the relation to the
variational principle in discrete space-time as introduced in~\cite{PFP}.
In Section~\ref{sec3} we prove Plancherel-type formulas for
Lorentz invariant functions. These Plancherel formulas are basic for the
subsequent reformulation of our variational principle in momentum space.
In Section~\ref{sec4} we outline the basic method and explain the
connection to state stability, whereas Sections~\ref{sec5} and~\ref{sec6}
are devoted to the more technical and computational aspects.
In Section~\ref{sec7} we introduce two additional free parameters, which
arise in the context of state stability, and built them into our framework
by introducing the so-called extended action.
In Section~\ref{sec8} we outline our numerical methods
and discuss a few examples of numerical minimizers and state stable Dirac sea
configurations.

\section{A Variational Principle for Dirac Sea Configurations}
\setcounter{equation}{0}  \label{sec2}
In this section we shall formulate a variational principle for Lorentz
invariant Dirac sea configurations. For clarity and self-consistency,
we proceed without referring to the variational principle in discrete
space-time as introduced in~\cite{PFP}. The connection to the latter variational principle
will be explained later in this section (see after Definition~\ref{def22}).
We begin with a general ansatz for the fermionic projector of the vacuum involving an
arbitrary number of generations~$g \in \N$ of Dirac seas of masses~$m_1, \ldots, m_g \in \R$,
which are taken into account with weight factors~$\rho_1, \ldots, \rho_g \geq 0$
(for a discussion of the weight factors see~\cite[Appendix~A]{F4}).
More precisely, we introduce the distribution in momentum space
\beq \label{hPdef}
\hat{P}(k) \;=\; \sum_{\beta=1}^g \rho_\beta\; (k\slsh+m_\beta)\, \delta(k^2-m_\beta^2)\,
\Theta(-k^0)
\eeq
and consider its Fourier transform to position space
\beq \label{fourier}
P(\xi) \;=\; \int \frac{d^4k}{(2 \pi)^4}\: \hat{P}(k)\: e^{i k \xi} \:.
\eeq
Here~$k \xi$ denotes the inner product of signature~$(+--\,-)$ in Minkowski space~$M$.

We first define our action formally and give the analytic justification
afterwards. For the upper and lower light cone and its boundary we use the notations
\[ I^\vee \;=\; \{ \xi \:|\: \xi^2 > 0, \,\xi^0>0 \}\:,\quad
I^\wedge \;=\; \{ \xi \:|\: \xi^2 > 0, \,\xi^0<0 \} \:,\quad
L \;=\; \{ \xi \:|\: \xi^2 = 0 \}\:. \]
For any~$\xi \in I^\vee$
we introduce the
so-called closed chain~$A$ and its trace-free part~$A_0$ by
\beq \label{Adef}
A \;=\; P(\xi)\, P(\xi)^*\:,\qquad A_0 \;=\; A - \frac{1}{4}\, \Tr(A) \:,
\eeq
where the star denotes the adjoint with respect to the indefinite inner
product~$\overline{\Psi} \Phi$ on the Dirac spinors (where~$\overline{\Psi}=
\Psi^\dagger \gamma^0$ is the adjoint spinor). Thus
\[ P(\xi)^* \;\equiv\; \gamma^0 P(\xi)^\dagger \gamma^0\:, \]
where the dagger denotes the transposed complex conjugate matrix.
Due to Lorentz invariance, $A_0$ can be written in the upper light cone as
\beq \label{Avee}
A_0 \;=\; \frac{\xi\slsh}{2}\:f(\xi^2) \qquad {\mbox{for~$\xi \in I^\vee$}}\:,
\eeq
and the relation~$A_0^*=A_0$ implies that the function~$f$ is real.
Hence the Lagrangian
\beq \label{Ldef}
{\mathcal{L}} \;=\; \Tr(A_0^2) \;=\; \xi^2\, f(\xi^2)^2
\eeq
is non-negative and depends only on the positive parameter~$z:=\xi^2$.
Thus the integral
\beq \label{Sdef}
{\mathcal{S}} \;\stackrel{\mbox{\scriptsize{formally}}}{=}\;
\int_0^\infty {\mathcal{L}}(z)\: z\, dz
\eeq
formally defines a positive functional depending on the free
parameters~$m_1, \ldots, m_g$ and $\rho_1, \ldots, \rho_g$.

Let us now give the definitions~(\ref{Adef}, \ref{Ldef}, \ref{Sdef}) a rigorous
mathematical meaning.
An explicit calculation of the Fourier transform~(\ref{fourier})
(see~\cite[Section~3]{F4}) shows that the distribution~$P(\xi)$
has the following properties. First, it is singular
on the light cone~$L$, but is a smooth function otherwise.
Furthermore, it is Lorentz invariant. This means in the upper light
cone that~$P(\xi)$ can be written as
\beq \label{Prep}
P(\xi) \;=\; \xi\slsh\, v(\xi^2)\:+\: h(\xi^2) \qquad {\mbox{for~$\xi \in I^\vee$}}
\eeq
with smooth complex functions~$v, h \in C^\infty(\R^+)$, which can be given
explicitly in terms of Bessel functions.
Thus the pointwise computations~(\ref{Adef}) and~(\ref{Ldef}) make sense and
\[ f(z) \;=\; {\mbox{Re}} \left(
v(z) \,\overline{h(z)} \right) \;\in\; C^\infty(\R^+)\:. \]
We conclude that the integrand in~(\ref{Sdef}) is a smooth function.
Near infinity, an asymptotic expansion of the Bessel functions
yields that the function~$f$ decays like~${\mathcal{O}}(z^{-2})$,
and thus the integral in~(\ref{Sdef}) is absolutely convergent at
infinity. Near the origin, $f$ has the expansion
(see~\cite[eq.~(3.5)]{F4})
\beq \label{fexp}
f(z) \;=\; \frac{\m_3}{z^2} + \frac{\m_5}{z} \:+\: {\mathcal{O}}(\log z)\:,
\eeq
where we set
\begin{eqnarray}
\m_3 &=& -\frac{1}{64\, \pi^5}
\sum_{\alpha, \beta=1}^g \rho_\alpha\, \rho_\beta\:(m_\alpha^3+m_\beta^3)
\label{m3def} \\
\m_5 &=& \frac{1}{512\, \pi^5} \sum_{\alpha, \beta=1}^g  \rho_\alpha\, \rho_\beta\:
(m_\alpha-m_\beta)^2\, (m_\alpha+m_\beta)^3 \:. \label{m5def}
\end{eqnarray}
Thus according to~(\ref{Ldef}),
the integrand in~(\ref{Sdef}) has a non-integrable pole at~$z=0$,
\beq \label{actreg}
\Tr(A_0^2) \, z \;=\; f(z)^2 \, z^2 \;=\;
\frac{\m_3^2}{z^2} + \frac{2\, \m_3 \,\m_5}{z} + {\mathcal{O}}(\log z)\:.
\eeq
In order to make sense of the integral, we need to subtract suitable counter terms.
The simplest method is to add indefinite integrals
of the poles in~(\ref{actreg}) evaluated at~$z=\varepsilon$,
\[ \lim_{\varepsilon \searrow 0} \left( \int_\varepsilon^\infty
\Tr(A_0^2)\: z\, dz \:-\: \frac{\m_3^2}{\varepsilon} \:+\:
2 \,\m_3\, \m_5\, \log \varepsilon \right) . \]
However, since the indefinite integrals are determined only up to
constants, it seems natural to allow for adding
arbitrary constants times~$\m_3$ and~$\m_3 \m_5$.
This freedom of modifying the action will turn out to be
very useful later. In order to be even more
flexible, it is preferable to treat the freedom
more generally by a function~$F(\m_3, \m_5)$.
This gives rise to the following definition.
\begin{Def} {\bf{(Lorentz invariant action)}} \label{def21}
For any given function~$F \in C^\infty(\R \times \R, \R)$ we define the
action~${\mathcal{S}}={\mathcal{S}}(m_1, \ldots, m_g, \rho_1, \ldots, \rho_g)$ by
\beq \label{Smdef}
{\mathcal{S}} \;=\;
\lim_{\varepsilon \searrow 0} \left( \int_\varepsilon^\infty
\Tr(A_0^2)\: z\, dz \:-\: \frac{\m_3^2}{\varepsilon} +
2 \,\m_3\, \m_5\, \log \varepsilon \right)
\;+\; F(\m_3, \m_5)\:.
\eeq
Here~$A_0$ is defined by~(\ref{Adef}) for any~$\xi \in I^\vee$
and~$z = \xi^2$. The parameters~$\m_3$ and $\m_5$ are
defined by~(\ref{m3def}, \ref{m5def}).
\end{Def}
Note that due to the negative counter terms, the action~(\ref{Smdef}) need no longer
be positive. But for any fixed~$\varepsilon>0$ it is clearly bounded from below.
We remark that in Definition~\ref{defextact} we shall extend this action by
additional summands involving two more free parameters.

Next we want to introduce a corresponding variational principle. In order to
avoid the trivial minimizers~$\rho_1=\cdots=\rho_g=0$ or~$m_1=\cdots=m_g=0$,
we need to impose a constraint.
\begin{Def} {\bf{(Lorentz invariant variational principle)}} \label{def22}
We minimize the action~${\mathcal{S}}$,  (\ref{Smdef}), varying the parameters~$\rho_1, \ldots, \rho_g \geq 0$
and~$m_1, \ldots, m_g \geq 0$ under the constraint
\beq \label{constraint}
{\mathcal{T}} := \sum_{\beta=1}^g \rho_\beta\: m_\beta^3 \;=\; 1\:.
\eeq
\end{Def}

In order to understand the relation between the above variational principle
and the variational principle introduced in discrete space-time in~\cite[\S3.5]{PFP}
and~\cite{F1}, we first consider the Lagrangian~(\ref{Ldef}) for
vectors~$\xi$ which are space-like or inside the
past light cone~$I^\wedge$.
Taking the adjoint of~(\ref{fourier}) and~(\ref{hPdef}), one sees that
\[ P(-\xi) \;=\; P(\xi)^* \:. \]
Hence the closed chain~(\ref{Adef}) can be written as~$A=P(\xi)\, P(-\xi)$.
Due to the ansatz~(\ref{hPdef}, \ref{fourier}), the matrices~$P(\pm \xi)$ are of the
form~$\alpha_\pm \xi\slsh + \beta_\pm$ with~$\alpha_\pm, \beta_\pm \in \C$.
This shows in particular that the matrices~$P(\xi)$ and~$P(-\xi)$ commute, and thus
\beq \label{Asymm}
A(-\xi) \;=\; A(\xi) \qquad {\mbox{and}} \qquad A_0(-\xi) \;=\; A_0(\xi) \:.
\eeq
For space-like~$\xi$, we can use Lorentz symmetry to write
the matrix~$A_0(\xi)$
in the form $A_0(\xi) = \xi\slsh g(\xi^2)$ with
a function~$g \in C^\infty((-\infty, 0))$. Replacing~$\xi$ by~$-\xi$
and using~(\ref{Asymm}), we see that~$g$ must vanish identically.
Combining this with~(\ref{Avee}) and~(\ref{Asymm}), we conclude
that away from the light cone, $A_0$ is of the form
\beq \label{tA}
A_0(\xi) \;=\; \frac{\xi \slsh}{2} \, f(\xi^2)\, \Theta(\xi^2)\, \epsilon(\xi^0)
 \qquad {\mbox{if~$\xi \not \in L$}}
\eeq
(where~$\epsilon$ is the step function~$\epsilon(x)=1$ if~$x \geq 0$
and~$\epsilon(x)=-1$ otherwise).
In words, $A_0$ is {\em{reflection symmetric}}~(\ref{Asymm}),
{\em{causal}} and {\em{Lorentz invariant}}.
Moreover, for timelike~$\xi$ the roots~$\lambda_1, \ldots, \lambda_4$
of the characteristic polynomial of~$A$ (counted with multiplicities) are
computed to be real. According to~\cite[Lemma~2.1]{F4}, these roots
all have the same sign. Combining these facts, we can write the
the Lagrangian~(\ref{Ldef}) in the alternative form
\[ {\mathcal{L}} \;=\; \Tr(A^2) \:-\: \frac{1}{4}\, \Tr(A)^2
\;=\; |A^2| - \frac{1}{4}\, |A|^2 \qquad {\mbox{if~$\xi^2 \neq 0$}} , \]
where~$|\,.\,|$ is the spectral weight (see~\cite[\S3.5]{PFP})
\[ |A| \;=\; \sum_{i=1}^{4} |\lambda_i|\:. \]
This is precisely the Lagrangian in the ``model example'' introduced in~\cite[\S3.5]{PFP},
for a particular value of the Lagrange multiplier.
The same Lagrangian also appears in~\cite{F1} as the so-called critical case
of the auxiliary variational principle.
As this Lagrangian vanishes for space-like~$\xi$ and is symmetric under
the transformation~$\xi \rightarrow -\xi$, it was no loss of generality
in Definition~\ref{def21} to restrict attention to $\xi$ in the upper
light cone. Hence the only differences between the above
variational principle and the variational principle in~\cite{PFP, F1} are the
following points:
\begin{description}
\item[(i)] A regularization on the light cone is used in~(\ref{Smdef}).
\item[(ii)] The additional constraint~(\ref{constraint}) appears.
\item[(iii)] The method of summing or integrating
over space-time is different. We can describe the
transition from the action in~\cite{PFP, F1} to~(\ref{Sdef}) by the
replacement
\[ \sum_{x, y \in M} \cdots \;\longrightarrow\; \int_0^\infty \cdots \;z\, dz\:. \]
\end{description}
Let us briefly discuss these points. It is not astounding that a regularization
on the light cone is necessary, because
the fermionic projector has poles and singularities on the light cone.
Such regularizations have been used previously in the
derivation of the continuum limit~\cite[Chapter~4]{PFP}
and for the vacuum fermionic projector in~\cite{F4}. The new and pleasant feature
of the regularization~{\bf{(i)}} is that it does not break Lorentz invariance.
The necessity of the constraint~{\bf{(ii)}} can be understood as replacing
the condition in discrete space-time that the number of particles is fixed.
However, the detailed form of the constraint~(\ref{constraint})
will not become clear before Section~\ref{sec4}.
The replacement~{\bf{(iii)}} requires a detailed explanation.
The natural method to describe the transition from discrete to a continuum space-time
is to replace sums by space-time integrals,
\[ \sum_{x, y \in M} \cdots \;\longrightarrow\; \int_M d^4 x \int_M d^4 y \cdots
\;=\; \int_M d^4 x \int_M d^4 \xi \cdots \:, \]
where in the last step we made a change of variables.
For homogeneous systems as considered here, where the fermionic projector
depends only on the difference~$y-x$, the integrand of the~$x$-integral is a
constant. Thus this integral merely gives an infinite constant, and it seems
natural to simply drop it. Thus it remains to discuss the replacement
\beq \label{replace}
\int_M \cdots \:d^4 \xi \;\longrightarrow\; \int_0^\infty \cdots \;z\, dz\:.
\eeq
This replacement has no simple justification; it changes the variational principle
substantially. The integral on the right has the nice property that it can be finite
for Lorentz invariant integrands, whereas the left integral is necessarily
infinite for Lorentz invariant integrands, because the integrals over the hyperbola~$\xi^2={\mbox{const}}$ diverge.
Furthermore, the integration measure~$z \,dz$ in~(\ref{replace}) can be motivated from the fact that
it has the same dimension of length four as the measure~$d^4 \xi$.
However, we again point out that the replacement~(\ref{replace}) changes the variational principle completely.
Thus the variational principle introduced here is not merely a ``continuum version''
of the variational principle in~\cite{PFP, F1}. It is a  different variational
principle. Nevertheless, as we shall see in Section~\ref{sec4}, there is a close connection
between minimizers of the variational principle of Definition~\ref{def22} and
the notion of state stability as introduced in~\cite[\S5.6]{PFP}.

We close this section with the derivation of the corresponding Euler-Lagrange (EL) equations.
We denote the first variation of the masses~$m_\beta$ and weights~$\rho_\beta$
by~$\delta m_\beta$ and~$\delta \rho_\beta$, respectively.
We let~$\delta P$ be the corresponding first variation of the
fermionic projector~(\ref{hPdef}, \ref{fourier}). The variation of the Lagrangian is
\[ \delta \, \Tr(A_0^2) \;=\; 2\, \Tr(A_0\, \delta A_0) \;=\; 2\, \Tr(A_0\, \delta A) \:, \]
where in the last step we used that the scalar component of~$\delta A$ drops out of the
trace because the other factor~$A_0$ only has a vector component. Using~(\ref{Adef})
and~(\ref{Asymm}), we conclude that
\[ \delta \, \Tr(A_0^2) \;=\; \Tr \Big( A_0 \left(\delta P(\xi)\: P(\xi)^* + P(\xi)\: \delta P(\xi)^*
\right) \Big) \;=\; 2 \,{\mbox{Re}} \, \Tr \Big( A_0 \: P(\xi)\: \delta P(-\xi) \Big) . \]
Treating the constraint with a Lagrange multiplier~$\lambda$, the EL equations become
\begin{eqnarray}
\lefteqn{ \lim_{\varepsilon \searrow 0} \left( \int_\varepsilon^\infty
{\mbox{Re}}\, \Tr(A_0\, P(\xi)\, \delta P(-\xi))\: z\, dz \:-\: \frac{\m_3\, \delta \m_3}{\varepsilon} +
(\delta \m_3\, \m_5 + \m_3\, \delta \m_5)\, \log \varepsilon \right) } \nonumber \\
&& + \delta F(\m_3, \m_5) - \lambda \, \delta {\mathcal{T}} \;=\; 0 \:. \spc\spc\spc\spc\spc\spc\spc
\label{ELp}
\end{eqnarray}
This equation looks rather complicated, and in this form it does not seem helpful for
understanding our variational principle. In order to clarify the situation, in the next sections
we shall transform our variational principle to momentum space.

\section{A Plancherel Formula for Lorentz Invariant Causal Functions}
\setcounter{equation}{0}  \label{sec3}
Studying the Lorentz invariant action~(\ref{Smdef}) directly in position space
has the following disadvantages. First, expressing~$A_0$ explicitly using Bessel function,
one sees that~$A_0(z)$ is oscillatory for large~$z$, making it difficult to
analytically estimate or to numerically compute the integral in~(\ref{Smdef}).
Second, the corresponding EL equations in position space~(\ref{ELp})
have no obvious interpretation.
For these reasons, it is preferable to transform the action~(\ref{Smdef}) to
momentum space. To this end, we need a Plancherel-type
formula, which relates the integral over the position variable~$z = \xi^2$
to an integral over a momentum variable~$a:=k^2$. In this section we shall
derive such Plancherel formulas.
For clarity, we begin with scalar functions. Thus, leaving out the Dirac matrices
in~(\ref{tA}), we consider functions of the form
\beq \label{Fdef}
F(\xi) \;=\; f(\xi^2)\, \Theta(\xi^2)\, \epsilon(\xi^0) \:,\qquad
G(\xi) \;=\; g(\xi^2)\, \Theta(\xi^2)\, \epsilon(\xi^0)
\eeq
with measurable complex-valued functions~$f$ and~$g$.
In view of the integration measure in~(\ref{Sdef}), we introduce on such functions the
inner product
\beq \label{FGip}
\langle F, G \rangle \;:=\; \int_0^\infty \overline{f(z)}\, g(z)\: z\, dz \:.
\eeq
We denote the space of functions for which the last integral converges absolutely,
by $L^2(M, z\, dz)$. For the upper and lower mass cone and their union
we introduce the notations
\[ {\mathcal{C}}^\vee \;=\; \{ k \:|\: k^2 > 0,\: k^0 > 0 \}\:, \quad
{\mathcal{C}}^\wedge \;=\; \{ k \:|\: k^2 > 0,\: k^0 < 0 \}\:, \quad
{\mathcal{C}} \;=\; \{ k \:|\: k^2 > 0 \}\:. \]
Taking the usual Fourier transform in Minkowski space,
\beq \label{fourier2}
\hat{F}(k) \;=\;  \int d^4 \xi \: F(\xi) \: e^{-i k \xi}\:,
\eeq
a symmetry argument shows that~~$\hat{F}(-k) = -F(k)$, and from
Lorentz invariance we conclude that~${\mbox{supp}} \hat{F} \subset \overline{\mathcal{C}}$.
Hence~$\hat{F}$ can be written similar to~(\ref{Fdef}) as
\beq \label{Fhdef}
\hat{F}(k) \;=\; \hat{f}(k^2)\, \Theta(k^2)\, \epsilon(k^0) \:.
\eeq
On such functions we introduce similar to~(\ref{FGip}) the inner product
\[ \langle \hat{F}, \hat{G} \rangle \;:=\; \frac{1}{(2 \pi)^4}
\int_0^\infty \overline{\hat{f}(a)}\, \hat{g}(a)\: a\, da\:, \]
where we set~$a = k^2$. The space of functions, for which the last integral converges absolutely, is denoted by~$L^2(\hat{M}, a\, da)$.
\begin{Thm} {\bf{(Lorentz invariant Plancherel formula, scalar case)}} \label{thmplancherel}
For functions of the form~(\ref{Fdef}), the Fourier transform~(\ref{fourier2}) is a unitary mapping from~$L^2(M, z\, dz)$ to $L^2(\hat{M}, a\, da)$. In particular,
for all~$F, G \in L^2(M, z\, dz)$,
\[ \int_0^\infty \overline{f(z)}\, g(z)\: z\, dz \;=\; \frac{1}{(2 \pi)^4}
\int_0^\infty \overline{\hat{f}(a)}\, \hat{g}(a)\: a\, da\:. \]
\end{Thm}
{\Proof} A possible method would be to express the transformation from~$f$ to~$\hat{f}$ 
with Bessel functions, and to use properties of the so-called Hankel transformation
(see for example~\cite{Z} and~\cite{W}). In order to avoid working with special functions,
we shall here use a different method, which also has the advantage of explaining
the integration measures~$z \,dz$ and~$a \,da$. Nevertheless, we briefly outline the
connection to the Hankel transformation: Rewriting the Fourier integral~(\ref{fourier2}) as
\[ \hat{F}(k) \;=\; \int_0^\infty f(z)\, dz \int d^4 \xi\: \delta(\xi^2-z)\: \epsilon(\xi^0)\:
e^{-i k \xi} \:, \]
we can carry out the last integral using Bessel functions. Comparing with~(\ref{Fhdef})
we obtain
\[ \hat{f}(a) \;=\; 2 i \pi^2 \int_0^\infty f(z) \: a \: \frac{J_1(\sqrt{a z})}{\sqrt{a z}}\: dz \:. \]
After transforming to the variables~$x=\sqrt{z}$ and~$y=\sqrt{a}$, we can apply
Parseval's equation for the Hankel transform~\cite[Theorem~5.1.2]{Z}.

We now give the alternative proof which avoids special functions.
Using a standard approximation argument, it suffices to prove the theorem for
functions~$F$ and~$G$ of the form~(\ref{Fdef}) with smooth and compactly supported
$f, g \in C^\infty_0(\R_+)$. A short calculation shows that then the wave operator becomes
\[ -\Box F(\xi) \;=\; (W f)(\xi^2)\: \Theta(\xi^2)\: \epsilon(\xi^0)\:, \]
where~$W$ is the ordinary differential operator
\beq \label{Adef2}
(W f)(z) \;=\; -\frac{4}{z}\, \frac{d}{dz} \left( z^2\: \frac{d}{dz}\, f \right) .
\eeq
Integrating by parts, one sees that~$W$ is a non-negative symmetric operator on the
Hilbert space~$L^2(\R^+, z\, dz)$ with domain of definition~$C^\infty_0(\R^+)$.

In order to construct a self-adjoint extension of~$W$, we use the well-known fact that
the operator~$-\partial_z^2$ is self-adjoint on~$L^2(\R^+, dz)$ with
domain~${\mathcal{D}}(-\partial_z^2) = H^{2,2}_0(\R^+)$. We set
\beq \label{domain}
{\mathcal{D}}(W) \;=\; \left\{ u {\mbox{ measurable with }} z \, u(z) \in
{\mathcal{D}}(-\partial_z^2) \right\} .
\eeq
To verify that~$W$ is self-adjoint, we use the definition of self-adjointness.
Suppose that for given~$v, w \in L^2(\R^+, z\, dz)$,
\beq \label{sarel}
\langle W u, v \rangle_{L^2(\R^+, z\, dz)} \;=\;
\langle u, w \rangle_{L^2(\R^+, z\, dz)} \qquad \forall u \in {\mathcal{D}}(W)\: .
\eeq
A short calculation shows that
\[ W u \;=\; -\partial_z^2 \tilde{u} \qquad {\mbox{with}} \qquad
\tilde{u}(z) \,:=\, z\, u(z)\:. \]
This allows us to rewrite~(\ref{sarel}) as
\[ \langle -\partial_z^2 \tilde{u}, \tilde{v} \rangle_{L^2(\R^+, dz)} \;=\;
\langle \tilde{u}, w \rangle_{L^2(\R^+, dz)} \qquad \forall \tilde{u} \in
{\mathcal{D}}(-\partial_z^2)\: . \]
Since the operator~$(-\partial_z^2)$ is self-adjoint, it follows that~$\tilde{v} \in
{\mathcal{D}}(-\partial_z^2)$
and that~$-\partial_z^2 \tilde{v} = w$. Hence~$v \in {\mathcal{D}}(W)$ and~$Wv=w$,
showing that~$W$ with the domain~(\ref{domain}) is indeed self-adjoint.

The spectral theorem yields
\beq \label{st}
\langle f, g \rangle_{L^2(\R^+, z\, dz)} \;=\; \int_{\sigma(A)} 
\langle f, dE_a  \,g \rangle_{L^2(\R^+, z\, dz)} \:,
\eeq
where~$dE_a$ is the corresponding spectral measure.
In order to get more information on the spectrum, it is most convenient to work
in momentum space, where the operator~$-\Box$, and consequently also
the operator~$W$, coincide with the multiplication operator~$\hat{f}(k) \mapsto
k^2 \hat{f}(k)$. Then it becomes obvious that~$\sigma(W)=\R^+ \cup \{0\}$
and that the spectral measure of~$W$ is absolutely continuous with respect to the
Lebesgue measure~$da$. Furthermore, the functional calculus can be expressed by
\[ \widehat{(h(W) f)}(b) \;=\;h(b) \,\hat{f}(b) \:, \]
implying that the spectral measure satisfies the relation
\beq \label{delrel}
\widehat{(dE_a f)}(b) \;=\; \delta(b-a)\, \hat{f}(b)\, da\:.
\eeq
As a consequence, the integrand in~(\ref{st}) can be written as
\beq \label{diagonal}
\langle f, dE_a  \,g \rangle_{L^2(\R^+, z\, dz)} \;=\;
\overline{\hat{f}(a)}\, \hat{g}(a)\; \rho(a)\, da
\eeq
with a non-negative measurable function~$\rho$.

In order to determine~$\rho$, we compute the left side of~(\ref{diagonal})
with the help of~(\ref{delrel}) to obtain
\begin{eqnarray*}
\overline{\hat{f}(a)}\, \hat{g}(a)\; \rho(a) &=& \int_0^\infty z\, dz
\int_M \frac{d^4k}{(2 \pi)^4}\: e^{- i \omega \sqrt{z}}\: \overline{\hat{f}(k^2)}\:
\Theta(k^2)\, \epsilon(k^0) \\
&& \qquad\quad \times \int_M \frac{d^4k'}{(2 \pi)^4}\: e^{i \omega' \sqrt{z}}\: \hat{g}(k'^2)\:
\delta(k'^2-a)\, \epsilon(k'^0) \:,
\end{eqnarray*}
where~$\omega$ and~$\omega'$ denote the zero components of~$k$ and~$k'$, respectively.
Using the scaling~$a \rightarrow \lambda^2 a$, $k \rightarrow \lambda k$, $k' \rightarrow
\lambda k'$ and~$z \rightarrow z/\lambda^2$, we find that
\[ \overline{\hat{f}(\lambda^2 a)}\, \hat{g}(\lambda^2 a)\; \rho(\lambda^2 a) \;=\;
\overline{\hat{f}(\lambda^2 a)}\, \hat{g}(\lambda^2 a)\; \lambda^2 \,\rho(a) \]
and thus~$\rho(a) = c\, a$ with a positive constant~$c$.
Using this fact in~(\ref{diagonal}), the abstract formula~(\ref{st}) becomes
\[ \int_0^\infty \overline{f(z)}\, g(z)\: z\, dz \;=\; c
\int_0^\infty \overline{\hat{f}(a)}\, \hat{g}(a)\: a\, da\:. \]
We finally determine~$c$ using the symmetry between position and momentum space
together with the fact that the definitions of the Fourier transformation and its
inverse differ by factors of~$(2 \pi)^4$ (compare~(\ref{fourier}) and~(\ref{fourier2})).
\QED

We now extend the last theorem to include Dirac matrices. We thus consider
similar to~(\ref{Fdef}) a function~$F$ (and analogously~$G$) of the form
\beq \label{Fdef2}
F(\xi) \;=\; \frac{\xi \slsh}{2}\, f(\xi^2)\, \Theta(\xi^2)\, \epsilon(\xi^0) \:.
\eeq
In the inner product we combine the two resulting factors~$\xi\slsh$ to a
factor~$z$,
\beq \label{ip1}
\langle F, G \rangle \;:=\; \int_0^\infty z \,\overline{f(z)}\, g(z)\: z\, dz \:.
\eeq
Then the Fourier transform~$\hat{F}(k)$, again defined by~(\ref{fourier2}), can be
written as
\beq \label{hFkdef}
\hat{F}(k) \;=\; \frac{k \slsh}{2} \,\hat{f}(k^2)\, \Theta(k^2)\, \epsilon(k^0) \:.
\eeq
The inner product in momentum space is defined by
\beq \label{ip2}
\langle \hat{F}, \hat{G} \rangle \;=\; \frac{1}{(2 \pi)^4}
\int_0^\infty a \,\overline{\hat{f}(a)}\, \hat{g}(a)\: a\, da  \:.
\eeq
With a slight abuse of notation, we denote the Hilbert spaces corresponding to
the inner products~(\ref{ip1}) and~(\ref{ip2}) again by~$L^2(M, z\, dz)$
and~$L^2(\hat{M}, a\, da)$, respectively.
\begin{Corollary} {\bf{(Lorentz invariant Plancherel formula, vector case)}} \label{corplan}
For functions of the form~(\ref{Fdef2}), 
the Fourier transform~(\ref{fourier2}) is a unitary mapping
from~$L^2(M, z\, dz)$ to $L^2(\hat{M}, a\, da)$. In particular, for all~$F, G \in L^2(M, z\, dz)$,
\[ \int_0^\infty z\, \overline{f(z)}\, g(z)\: z\, dz \;=\; \frac{1}{(2 \pi)^4}
\int_0^\infty a\, \overline{\hat{f}(a)}\, \hat{g}(a)\: a\, da\:. \]
\end{Corollary}
{\Proof} We can again assume that~$f, g \in C^\infty_0(\R_+)$. Introducing the scalar function
\[ F_s(\xi) \;=\;  f_s(\xi^2)\, \Theta(\xi^2)\, \epsilon(\xi^0) \quad {\mbox{with}} \quad
f_s \equiv f\:, \]
we obviously have~$F(\xi) = \xi\slsh F_s(\xi)/2$. Since multiplication in position space
corresponds to differentiation in momentum space, we know that
\[ \hat{F}(k) \;=\; \frac{i \Pdd_k}{2} \,\hat{F}_s(k)\:. \]
Representing~$\hat{F}_s$ in the form~(\ref{Fhdef}) with an overall subscript~$s$,
it follows that the function~$\hat{f}$ in~(\ref{hFkdef}) is given by
\[ \hat{f}(a) \;=\; 2 \,\hat{f}'_s(a) \:. \]
Finally, we know that~$\widehat{\xi^2 F_s(\xi)} = -\Box_k \hat{F}_s(k)$ and thus
\[ \widehat{z f_s(z)} \;=\; \hat{W} \hat{f}_s(a)\:, \]
where~$\hat{W}$ is minus the wave operator in momentum space, i.e.\ in analogy to~(\ref{Adef2})
\beq \label{woms}
(\hat{W} \hat{f}_s)(a) \;=\; -\frac{4}{a}\, \frac{d}{da} \left( a^2\: \frac{d}{da}\,
\hat{f}_s \right) .
\eeq
Using for~$G$ the same notation as for~$F$, we obtain
\begin{eqnarray*}
\lefteqn{ \int_0^\infty z\, \overline{f(z)}\, g(z)\: z\, dz \;=\;
\int_0^\infty \overline{z \,f_s(z)}\, g_s(z)\: z\, dz } \\
&\stackrel{(*)}{=}& \frac{1}{(2 \pi)^4} \int_0^\infty \overline{\widehat{z f_s}(a)}\, \hat{g}_s(a)\: a\, da
\;=\; \frac{1}{(2 \pi)^4} \int_0^\infty \left(\hat{W} \overline{\hat{f}_s(a)}\right)
\hat{g}_s(a)\: a\, da \\
&\stackrel{(+)}{=}& \frac{1}{(2 \pi)^4} \int_0^\infty 4 \,\overline{\hat{f}_s'(a)}\, \hat{g}_s'(a)\: a^2\, da
\;=\; \frac{1}{(2 \pi)^4} \int_0^\infty a\, \overline{\hat{f}(a)}\, \hat{g}(a)\: a\, da\,,
\end{eqnarray*}
where in~(*) we applied Theorem~\ref{thmplancherel}, and in~(+) we used the
explicit form of~$\hat{W}$, (\ref{woms}), and integrated by parts.
\QED

\section{The Variational Principle in Momentum Space, \\
Connection to State Stability}
\setcounter{equation}{0}  \label{sec4}
Using the Lorentz invariant Plancherel formula of Corollary~\ref{corplan}, we shall now
transform our action and the corresponding EL equations to momentum space. The main
difficulty is to treat the singularities of~$A_0$ on the light cone. We here merely describe the
regularization schematically; the detailed constructions involving a rather subtle
regularization technique will be carried out in the subsequent
Sections~\ref{sec5} and~\ref{sec6}.

We want to apply Corollary~\ref{corplan} to the integral in~(\ref{Smdef}).
The simplest approach would be to regularize~$A_0$ by multiplication with a
characteristic function,
\beq \label{A0reg}
A_0^\varepsilon(\xi) \;:=\; \Theta(\xi^2-\varepsilon)\: A_0(\xi)\:.
\eeq
According to~(\ref{tA}),
\[ A_0^\varepsilon(\xi) \;=\; \frac{\xi \slsh}{2} \, f^\varepsilon(\xi^2)\, \Theta(\xi^2)\,
\epsilon(\xi^0) \qquad {\mbox{with}} \qquad
f^\varepsilon(z) \;=\; \Theta(z-\varepsilon)\, f(z)\:. \]
Then
\begin{eqnarray}
\int_\varepsilon^\infty \Tr \left( A_0^2 \right) z\, dz
&=& \int_0^\infty \Tr \left( (A_0^\varepsilon)^2 \right) z\, dz \nonumber \\
&=& \int_0^\infty f^\varepsilon(z)^2\: z^2\, dz
\;=\; \frac{1}{(2 \pi)^4}  \int_0^\infty |\widehat{f^\varepsilon}(a)|^2
\: a^2\, da \:, \label{f2} 
\end{eqnarray}
where in the last step we applied Corollary~\ref{corplan}. Here~$\widehat{f^\varepsilon}$
is defined via the relation
\beq \label{f3}
\int d^4 \xi\: \Theta(\xi^2-\varepsilon)\: A_0\: e^{-ik\xi} \;=\;
\frac{k\slsh}{2}\; \widehat{f^\varepsilon}(k^2)\: \Theta(k^2)\, \epsilon(k^0)\:.
\eeq
The problem is that in~(\ref{f3}) we take the Fourier transform of a function
which is discontinuous on the hypersurface~$\{\xi \:|\: \xi^2 = \varepsilon \}$.
As a consequence, the function~$\widehat{f^\varepsilon}$ depends on the regularization
in a complicated oscillatory way, making it difficult to analyze the $\varepsilon$-dependence
of the last integral in~(\ref{f2}).

\begin{figure}[tb]
\begin{center}
\begin{picture}(0,0)%
\includegraphics{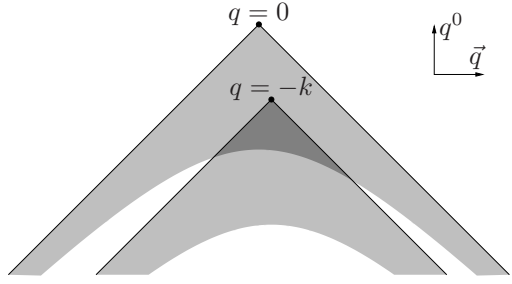}%
\end{picture}%
\setlength{\unitlength}{1380sp}%
\begingroup\makeatletter\ifx\SetFigFont\undefined%
\gdef\SetFigFont#1#2#3#4#5{%
  \reset@font\fontsize{#1}{#2pt}%
  \fontfamily{#3}\fontseries{#4}\fontshape{#5}%
  \selectfont}%
\fi\endgroup%
\begin{picture}(9044,4999)(-4521,-3683)
\put(  1,929){\makebox(0,0)[b]{\smash{{\SetFigFont{10}{12.0}{\rmdefault}{\mddefault}{\updefault}{$q=0$}%
}}}}
\put(226,-421){\makebox(0,0)[b]{\smash{{\SetFigFont{10}{12.0}{\rmdefault}{\mddefault}{\updefault}{$q=-k$}%
}}}}
\put(3241,659){\makebox(0,0)[lb]{\smash{{\SetFigFont{10}{12.0}{\rmdefault}{\mddefault}{\updefault}{$q^0$}%
}}}}
\put(3781,119){\makebox(0,0)[lb]{\smash{{\SetFigFont{10}{12.0}{\rmdefault}{\mddefault}{\updefault}{$\vec{q}$}%
}}}}
\end{picture}%
\caption{The convolution integral in~(\ref{cFdef}).}
\label{figconv1}
\end{center}
\end{figure}

To bypass this difficulty, we will use a different regularization method. First, we
extend~$A_0$ in the distributional sense across the light cone. More precisely,
we introduce~$\tM(\xi)$ as in~\cite{F4} by the requirements that
\beq \label{Mtdef}
\tM \in {\mathcal{S}}'(M) \qquad {\mbox{and}} \qquad
\tM(\xi) \;=\; 2 A_0(\xi) \quad \forall \xi \not \in L \:,
\eeq
where~${\mathcal{S}}'(M)$ denotes the space of tempered distributions.
Furthermore, $\tM$ should respect the symmetry~(\ref{Asymm}),
\beq
\tM(-\xi) \;=\; \tM(\xi)\:. \label{tMsymm}
\eeq
The Fourier transform of~$\tM$ is denoted by~$\hM \in {\mathcal{S}}'(\hat{M})$.
Using again the symmetry argument after~(\ref{fourier2}), we find that
\beq \label{hMsymm}
{\mbox{supp}}\, \hM \subset \overline{\mathcal{C}} \qquad {\mbox{and}} \qquad
\tM(-k) \;=\; \tM(k)\:.
\eeq
Next, we regularize~$\tM$ in a Lorentz invariant way which is simple and explicit
both in position and momentum space. The regularized~$\tM$ is
in~$L^2(M, z\, dz)$, making it possible to apply Corollary~\ref{corplan}.
This allows us to rewrite the action~(\ref{Smdef}) as
\beq \label{regact}
{\mathcal{S}} \;=\; \frac{1}{(2 \pi)^4} \:\lim_{\varepsilon \searrow 0} \left(
\int_0^\infty \frac{1}{4}\: \Tr \left( \hM^{\varepsilon}(k)^2 \right)
a\: da \;+\: F_{\varepsilon}(\m_3, \m_5) \right) ,
\eeq
where~$k \in \hat{M}$ is any vector with~$k^2=a$. Here
the new function~$F_{\varepsilon}$ also involves the counter terms. For the first
variation we then obtain
\beq \label{dS}
\delta {\mathcal{S}} \;=\; \frac{1}{(2 \pi)^4} \:\lim_{\varepsilon \searrow 0} \left(
\int_0^\infty \frac{1}{2}\: \Tr \left( \hM^\varepsilon \: \delta \hM^\varepsilon \right)
a\: da \;+\: \delta F_{\varepsilon}(\m_3, \m_5) \right) .
\eeq
According to~(\ref{Mtdef}) and~(\ref{Adef}), the distribution~$\tM(\xi)/2$ coincides
away from the light cone with the trace-free part of the product~$P(\xi)$
times~$P(-\xi)$. Since multiplication in position space corresponds to convolution
in momentum space, one might expect that~$\hM$ equals the convolution of~$\hat{P}$
by itself. More precisely, the change of variables
\begin{eqnarray*}
P(\xi)\, P(-\xi) &=& \int \frac{d^4p}{(2 \pi)^4}  \int \frac{d^4q}{(2 \pi)^4} 
\hat{P}(p)\, \hat{P}(q)\; e^{i (p-q) \xi} \\
&=& \int \frac{d^4k}{(2 \pi)^4} \left( \int \frac{d^4q}{(2 \pi)^4} 
\hat{P}(k+q)\, \hat{P}(q) \right) e^{i (p-q) \xi}
\end{eqnarray*}
suggests that it should be helpful to decompose~$\hM^\varepsilon$ as
\beq
\hM^\varepsilon(k) \;=\; \hF(k) \:+\:\hN^\varepsilon(k) \:, \label{hMdistr}
\eeq
where we set
\beq \label{cFdef}
\hF(k) \;:=\; 2 \int \frac{d^4q}{(2 \pi)^4} \:\left( \hat{P}(k+q)\: \hat{P}(q) \right)_0 \:,
\qquad {\mbox{defined if~$k^2>0$}}.
\eeq
Here the subscript zero means that we take only the vector component,
and the new function~$\hN^\varepsilon$ involves the dependence on the regularization.
As the poles of~$\hM$ are characterized by~$\m_3$ and~$\m_5$ (see~(\ref{fexp})),
it is not astonishing that~$\hN^\varepsilon$ will depend only on the parameters~$\m_3$,
$\m_5$ and~$\varepsilon$.
Using that~$\hat{P}$ is supported on the lower mass shells (see~(\ref{hPdef})),
one sees that the integrand in~(\ref{cFdef}) is compactly supported
(see Figure~\ref{figconv1} in the case when~$k \in {\mathcal{C}}^\vee$; the case~$k \in {\mathcal{C}}^\wedge$ is similar). 
As a consequence, the integral in~(\ref{cFdef}) will be finite.
Since~$\hM^\varepsilon \rightarrow \hM$ as~$\varepsilon \searrow 0$,
it is obvious from~(\ref{hMdistr}) that~$\hN^\varepsilon$ converges in this limit
to a function~$\hN(k)$ defined if~$k^2>0$. Using~(\ref{hMdistr}) in~(\ref{dS}), we obtain
\begin{eqnarray}
\delta {\mathcal{S}} &=& \frac{1}{(2 \pi)^4}\: \lim_{\varepsilon \searrow 0}
\int_0^\infty  a\: da \;\frac{1}{2}\:
 \Tr \left( \hM^\varepsilon(k) \: \delta \hF(k) \right) \label{dS1} \\
 &&+ \frac{1}{(2 \pi)^4}\: \lim_{\varepsilon \searrow 0} \left(
\int_0^\infty \frac{1}{2}\: \Tr \left( \hM^\varepsilon \: \delta \hN^\varepsilon
\right) a\: da \;+\: \delta F_{\varepsilon}(\m_3, \m_5) \right) . \label{dS2}
\end{eqnarray}
We shall see that in~(\ref{dS1}) the regularization can be removed. In~(\ref{dS2}), on the other hand, the integral diverges as~$\varepsilon \searrow 0$, but this divergence is
compensated by the counter terms in~$\delta F_\varepsilon$. We thus obtain the
following result, where for notational convenience we omit the irrelevant
prefactors~$(2 \pi)^{-4}$.
\begin{Lemma} \label{lemmalater}
The first variation of the action~(\ref{Smdef}) can be written as
\beq \label{dSrel0}
\delta {\mathcal{S}} \;=\;  \int_0^\infty  a\: da \int \frac{d^4q}{(2 \pi)^4}\:
 \Tr \left( \hM(k) \: \delta\!\left( \hat{P}(k+q)\: \hat{P}(q) \right)_0 \right) \:+\:
 \delta F(\m_3, \m_5)\:,
\eeq
where~$k \in \hat{M}$ is any vector with~$k^2=a$, and~$\delta F$ denotes the
variation of the function~$F$ in Definition~\ref{def21}.
\end{Lemma}
This concludes our sketch of the regularization procedure. This sketch will be our
guideline for the constructions in Sections~\ref{sec5} and~\ref{sec6}.
The proof of Lemma~\ref{lemmalater} will be given at the end of Section~\ref{sec6}.

We next consider the variation of the product in~(\ref{dSrel0}). Changing variables
\[ \int d^4q \: (\delta \hat{P}(k+q))\:  \hat{P}(q) \;=\;
\int d^4q \: (\delta \hat{P}(q))\:  \hat{P}(-k+q) \]
and using that commuting the two factors in the integrand does not change their vector
component, we find that varying the first factor is the same as varying the second
factor if we flip the sign of~$k$. In view of~(\ref{hMsymm}) we can thus write the
variation of the action as
\begin{eqnarray}
\delta {\mathcal{S}} &=& \int_0^\infty  a\: da \int \frac{d^4q}{(2 \pi)^4}\:
 \Tr \left( \left[ \hM(k) \: \hat{P}(k+q) \:+\: \hM(-k) \: \hat{P}(-k+q) \right]
 \: \delta \hat{P}(q) \right) \nonumber \\
&&+  \delta F(\m_3, \m_5)\:, \label{dSrel}
\end{eqnarray}
where~$k \in \hat{M}$ is an arbitrary vector with~$k^2=a$.
Note that we here omitted the subscript zero. This does not change the value of
the trace, because the factor~$\hM$ only has a vector component. 

\begin{figure}[tb]
\begin{center}
\begin{picture}(0,0)%
\includegraphics{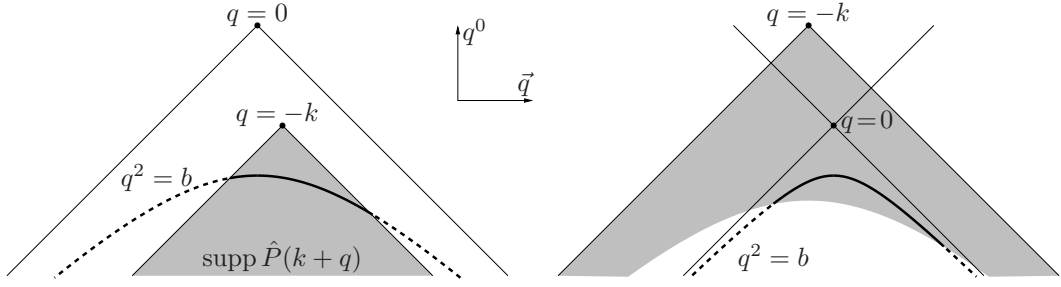}%
\end{picture}%
\setlength{\unitlength}{1380sp}%
\begingroup\makeatletter\ifx\SetFigFont\undefined%
\gdef\SetFigFont#1#2#3#4#5{%
  \reset@font\fontsize{#1}{#2pt}%
  \fontfamily{#3}\fontseries{#4}\fontshape{#5}%
  \selectfont}%
\fi\endgroup%
\begin{picture}(18944,5057)(-4521,-3741)
\put(  1,929){\makebox(0,0)[b]{\smash{{\SetFigFont{10}{12.0}{\rmdefault}{\mddefault}{\updefault}{$q=0$}%
}}}}
\put(361,-871){\makebox(0,0)[b]{\smash{{\SetFigFont{10}{12.0}{\rmdefault}{\mddefault}{\updefault}{$q=-k$}%
}}}}
\put(-1169,-2041){\makebox(0,0)[rb]{\smash{{\SetFigFont{10}{12.0}{\rmdefault}{\mddefault}{\updefault}{$q^2=b$}%
}}}}
\put(3691,569){\makebox(0,0)[lb]{\smash{{\SetFigFont{10}{12.0}{\rmdefault}{\mddefault}{\updefault}{$q^0$}%
}}}}
\put(4681,-331){\makebox(0,0)[lb]{\smash{{\SetFigFont{10}{12.0}{\rmdefault}{\mddefault}{\updefault}{$\vec{q}$}%
}}}}
\put(451,-3481){\makebox(0,0)[b]{\smash{{\SetFigFont{10}{12.0}{\rmdefault}{\mddefault}{\updefault}{$\mathrm{supp}\,\hat{P}(k+q)$}%
}}}}
\put(9901,929){\makebox(0,0)[b]{\smash{{\SetFigFont{10}{12.0}{\rmdefault}{\mddefault}{\updefault}{$q=-k$}%
}}}}
\put(9271,-3481){\makebox(0,0)[b]{\smash{{\SetFigFont{10}{12.0}{\rmdefault}{\mddefault}{\updefault}{$q^2=b$}%
}}}}
\put(10486,-1006){\makebox(0,0)[lb]{\smash{{\SetFigFont{10}{12.0}{\rmdefault}{\mddefault}{\updefault}{$q\!=\!0$}%
}}}}
\end{picture}%
\caption{The $q$-integral in~(\ref{qint}) in the cases~$k \in {\mathcal{C}}^\vee$
(left) and~$k \in {\mathcal{C}}^\wedge$ (right).}
\label{figconv2}
\end{center}
\end{figure}

\begin{figure}[tb]
\begin{center}
\begin{picture}(0,0)%
\includegraphics{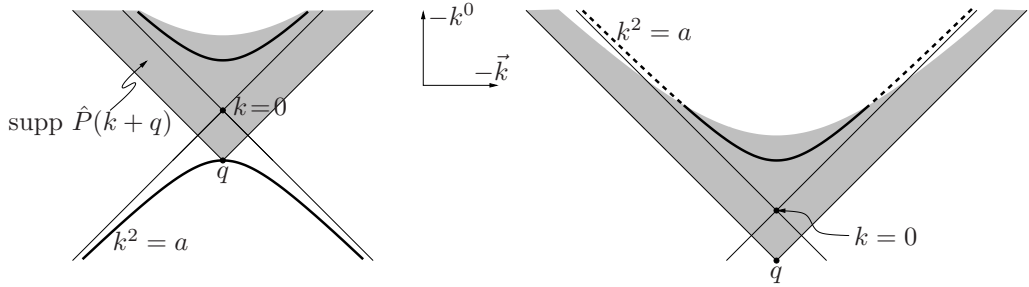}%
\end{picture}%
\setlength{\unitlength}{1380sp}%
\begingroup\makeatletter\ifx\SetFigFont\undefined%
\gdef\SetFigFont#1#2#3#4#5{%
  \reset@font\fontsize{#1}{#2pt}%
  \fontfamily{#3}\fontseries{#4}\fontshape{#5}%
  \selectfont}%
\fi\endgroup%
\begin{picture}(17189,5038)(-2721,-4130)
\put(3691,569){\makebox(0,0)[lb]{\smash{{\SetFigFont{10}{12.0}{\rmdefault}{\mddefault}{\updefault}{$-k^0$}%
}}}}
\put(4501,-421){\makebox(0,0)[lb]{\smash{{\SetFigFont{10}{12.0}{\rmdefault}{\mddefault}{\updefault}{$-\vec{k}$}%
}}}}
\put(-899,-1321){\makebox(0,0)[rb]{\smash{{\SetFigFont{10}{12.0}{\rmdefault}{\mddefault}{\updefault}{supp $\hat{P}(k+q)$}%
}}}}
\put(7066,299){\makebox(0,0)[lb]{\smash{{\SetFigFont{10}{12.0}{\rmdefault}{\mddefault}{\updefault}{$k^2=a$}%
}}}}
\put(9946,-3976){\makebox(0,0)[b]{\smash{{\SetFigFont{10}{12.0}{\rmdefault}{\mddefault}{\updefault}{$q$}%
}}}}
\put(181,-1051){\makebox(0,0)[lb]{\smash{{\SetFigFont{10}{12.0}{\rmdefault}{\mddefault}{\updefault}{$k\!=\!0$}%
}}}}
\put(-1979,-3481){\makebox(0,0)[lb]{\smash{{\SetFigFont{10}{12.0}{\rmdefault}{\mddefault}{\updefault}{$k^2=a$}%
}}}}
\put(  1,-2176){\makebox(0,0)[b]{\smash{{\SetFigFont{10}{12.0}{\rmdefault}{\mddefault}{\updefault}{$q$}%
}}}}
\put(11341,-3391){\makebox(0,0)[lb]{\smash{{\SetFigFont{10}{12.0}{\rmdefault}{\mddefault}{\updefault}{$k=0$}%
}}}}
\end{picture}%
\caption{The corresponding $k$-integrals in the two cases of Figure~\ref{figconv2}.}
\label{figconv3}
\end{center}
\end{figure}

The following geometric argument is crucial for getting an understanding of the EL equations.
It suffices to consider the first summand in the square bracket in~(\ref{dSrel}), because
the second term is then obtained by flipping the sign of~$k$. Using the support
property in~(\ref{hMsymm}), we may insert the integral over a $\delta$-distribution,
\begin{eqnarray}
\lefteqn{ \int_0^\infty a\, da \int d^4q\: \hM(k) \: \hat{P}(k+q)\: \delta \hat{P}(q) } \nonumber \\
&=& \int_0^\infty a\, da \int_0^\infty db \int d^4q\; \delta(q^2-b)\: 
\hM(k) \: \hat{P}(k+q)\: \delta \hat{P}(q)\:. \label{qint}
\end{eqnarray}
For fixed~$a$, $b$ and~$k \in \C$, the $q$-integral extends over a compact subset
of the hyperbola~$\{q \:|\: q^2=b\}$ (see Figure~\ref{figconv2}).
Any two vectors on this hyperbola can be transformed into each other by a Lorentz boost.
Thus, using Lorentz symmetry, we can assume that the vector~$q$ is fixed. Instead, the
vector~$k$ will no longer be constant, but it will run over a compact
subset of the hyperbola~$\{k \:|\: k^2=a\}$ (see Figure~\ref{figconv3}).
Taking into account the relative sizes of the $q$-subset and the corresponding $k$-subset,
we obtain the transformation rule
\[ \int_0^\infty a\, da \int d^4q \cdots \;=\; \int_0^\infty b\, db
\int_{{\mathcal{C}}^\wedge {\mbox{\scriptsize{ or }}} {\mathcal{C}}^\vee} d^4k \cdots \:, \]
where on the right we choose any vector~$q \in {\mathcal{C}}^\wedge$ with~$q^2=b$,
and the $k$-integral goes over~${\mathcal{C}}^\vee$ or~${\mathcal{C}}^\wedge$, depending on whether we started
with a vector~$k$ on the lower or upper mass cone.
Combining the contributions of the two summands in the square bracket in~(\ref{dSrel}),
we obtain a $k$-integral over~${\mathcal{C}}$. Finally, the support property in~(\ref{hMsymm})
allows us to extend the $k$-integration to all of~$\hat{M}$. We thus obtain
\[ \delta {\mathcal{S}} \;=\;  \int_0^\infty  b\: db \int \frac{d^4k}{(2 \pi)^4}\:
 \Tr \left( \hM(k) \: \hat{P}(k+q)  \: \delta \hat{P}(q) \right)
\:+\:  \delta F(\m_3, \m_5)  \:. \]
Next, it is convenient to transform the $k$-integral to a convolution integral,
\[ \int \frac{d^4k}{(2 \pi)^4}\: \hM(k) \: \hat{P}(k+q) \;=\;
\int \frac{d^4k}{(2 \pi)^4}\: \hM(k) \: \hat{P}(q-k)
\;=:\;  (\hM * \hat{P})(q) \:, \]
where we flipped the sign of ~$k$ and used the symmetry property of~$\hM$
in~(\ref{hMsymm}). Then we can write the variation of the action in the more compact form
\beq \label{dSrel1}
\delta {\mathcal{S}} \;=\;  \int_0^\infty
 \Tr \left( (\hM*\hat{P})(q)  \: \delta \hat{P}(q) \right)  b\: db
\:+\:  \delta F(\m_3, \m_5)  \:.
\eeq

Our last step is to treat the variation of the function~$F$. According to
the chain rule,
\beq \label{freec}
\delta F(\m_3, \m_5) \;=\; D_1 F(\m_3, \m_5)\, \delta \m_3 + D_2 F(\m_3, \m_5)\, \delta \m_5 \:.
\eeq
It is an important observation that
the relations~(\ref{Mtdef}) determine~$\tM$ only up to singular contributions on
the light cone. More specifically, there is the natural freedom to modify~$\tM(\xi)$ by
$\delta'$- and $\delta$-contributions of the form
\beq \label{tMmod}
\tM(\xi) \;\asymp\;
c_0\: \xi\slsh\, \delta'(\xi^2)\: \epsilon(\xi^0) \:+\:
c_1\: \xi\slsh\, \delta(\xi^2)\: \epsilon(\xi^0) \:.
\eeq
As we shall see later (see Lemma~\ref{lemmac0c1comp}),
by modifying~$c_0$ and~$c_1$ appropriately, we can precisely
absorb the term~(\ref{freec}) which arises in the variation of~$F$.
Conversely, the freedom to choose the function~$F$ in~(\ref{Smdef})
can be regarded as taking into account
the fact that the distribution~$\tM$ is determined only modulo
the contribution~(\ref{tMmod}). In what follows, we shall always
omit the term~$\delta F(\m_3, \m_5)$ in~(\ref{dSrel1}), giving the following result.

\begin{figure}[tb]
\begin{center}
\scalebox{0.9}
{\includegraphics{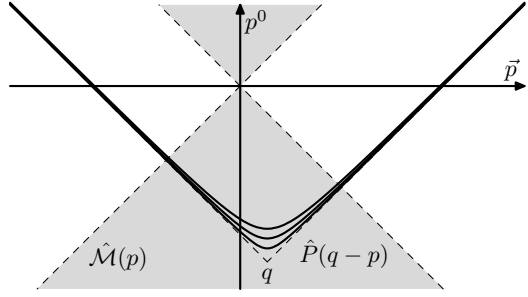}}
\caption{The convolution~${\hat{\mathcal{M}}} * \hat{P}$.}
\label{figMP}
\end{center}
\end{figure}

\begin{Prp} \label{prp41}
The first variation of the action~${\mathcal{S}}$, (\ref{Smdef}), can be written in
momentum space as
\beq \label{dSrel3}
\delta {\mathcal{S}} \;=\; 2 \int_0^\infty  
 \Tr \left( \hat{Q}(q)  \: \delta \hat{P}(q) \right) b\: db \:,
\eeq
where~$q \in {\mathcal{C}}^\wedge$ is any vector in the lower mass cone with~$q^2=b$ and
\beq \label{hQdef}
\hat{Q}(q) \;=\; \frac{1}{2}\: (\hM * \hat{P})(q) \;:=\;
\frac{1}{2} \int \frac{d^4p}{(2 \pi)^4}\: \hM(p)\, \hat{P}(q-p)\:.
\eeq
Here~$\hM$ is the Fourier transform of the distribution~$\tM$ as introduced by~(\ref{Mtdef}),
for specific values of the free parameters~$c_0$ and~$c_1$ in~(\ref{tMmod}).
\end{Prp}
To fully understand~(\ref{dSrel3}, \ref{hQdef}), one should observe that,
according to~(\ref{hPdef}), the factor~$\delta \hat{P}(q)$ in~(\ref{dSrel3})
is supported in~${\mathcal{C}}^\wedge$. But for~$q$ in the lower mass cone,
the convolution integral in~(\ref{hQdef}) is finite, because the integrand is
compactly supported (see Figure~\ref{figMP}).
The Fourier transform of~$\hat{Q}$ is simply the product
\beq \label{tMP}
Q(\xi) \;=\; \frac{1}{2}\: \tM(\xi)\: P(\xi)\:.
\eeq
The distribution~$Q(y-x)$ is precisely the kernel of the operator~$Q$ which
appears in the EL equations~$[P,Q]=0$ corresponding to the
variational principle introduced in~\cite[\S3.5]{PFP}.

The notion of state stability imposes conditions on the eigenvalues of
the matrix~$\hat{Q}(q)$. For the sake of self-consistency we here
restate the definition as given in~\cite[Def.~5.6.2]{PFP}.
\begin{Def} \label{def611}\index{state stability}
  The fermionic projector of the vacuum is called {\bf{state stable}}
  if the corresponding operator $\hat{Q}(k)$ is well-defined in the
  lower mass cone ${\mathcal{C}}^\land$ and can be written as
\beq \label{62f}
\hat{Q} (k) \;=\; a\:\frac{k\slsh}{|k|} + b \eeq
(where~$|k| \equiv \sqrt{k^2}$) with
continuous real functions $a$ and $b$ on $\mathcal{C}^\land$ having
the following properties:
\begin{description}
\item[(i)] $a$ and $b$ are Lorentz invariant,
\[ a \;=\; a(k^2)\:,\spc b \;=\; b(k^2) \:. \]
\item[(ii)] $a$ is non-negative.
\item[(iii)] The function $a+b$ is minimal on the mass shells,
\beq \label{(iii)}
(a+b)(m^2_\alpha) \;=\; \inf_{q \in {\mathcal{C}}^\land} (a+b)(q^2)
\qquad\mbox{$\forall \alpha \in \{1,\ldots, g\}$}.
\eeq
\end{description}
\end{Def}
In our setting, the representation~(\ref{62f}) as well as~(i) are obvious,
and thus we need to analyze~(ii) and~(iii).
The next theorem shows that the EL equations corresponding to the Lorentz invariant
variational principle of Definition~\ref{def22} give a necessary condition for
state stability. Our proof will also explain the detailed form of the constraint~(\ref{constraint}).
\begin{Thm} \label{thm1}
The fermionic projector is a critical point of the variational principle of Definition~\ref{def22}
if and only if the functions~$a$ and~$b$ in the representation~(\ref{62f}) have the
following properties:
\begin{description}
\item[(a)] $(a+b)(m_\alpha^2) \;=\; (a+b)(m_\beta^2) \qquad \forall \alpha,
\beta\in \{1,\ldots, g\}$
\item[(b)] $(a+b)'(m_\alpha^2) =\; 0 \qquad\qquad\qquad\;\; \forall \alpha \in \{1,\ldots, g\}\,.$
\end{description}
\end{Thm}
{\Proof} Using the ansatz~(\ref{hPdef}), we find
\beq \label{v1}
\frac{1}{2} \int_0^\infty \Tr \left( \hat{Q}(q)  \: \hat{P}(q) \right) b\: db 
\;=\; \frac{1}{2} \sum_{\beta=1}^g \rho_\beta\: m_\beta^3\; \Tr \left( \hat{Q}(q_\beta)\:
\frac{q \slsh_\beta + m_\beta}{m_\beta} \right) ,
\eeq
where~$q, q_\beta \in {\mathcal{C}}^\wedge$ are any vectors with~$q^2=b$ and~$q_\beta^2=m_\beta^2$.
Using~(\ref{62f}), this simplifies to
\beq \label{v2}
\frac{1}{2} \int_0^\infty \Tr \left( \hat{Q}(q)  \: \hat{P}(q) \right) b\: db 
\;=\; 2 \sum_{\beta=1}^g \rho_\beta\: m_\beta^3\; (a+b)(m_\beta^2)\:.
\eeq
According to Proposition~\ref{prp41}, $\delta {\mathcal{S}}$ is obtained from~(\ref{v1}) by
varying~$\rho_\beta$ and~$m_\beta$, keeping the operator~$\hat{Q}$ fixed.
Equivalently, we can vary~(\ref{v2}) for fixed functions~$a$ and~$b$.
We treat the constraint~(\ref{constraint}) as in~(\ref{ELp}) with a Lagrange
multiplier~$\lambda$. Varying the weights~$\rho_\beta$ gives the conditions
\beq \label{rvar}
2\, m_\beta^3\, (a+b)(m_\beta^2) \;=\; \lambda m_\beta^3 \qquad
\forall \beta \in \{1, \ldots, g\}\:,
\eeq
proving~(a). Similarly, varying the masses~$m_\beta$ and using~(\ref{rvar}) gives~(b).
\QED
The properties~(a) and~(b) in this theorem are weaker than the condition~(iii)
in the definition of state stability, because~(\ref{(iii)}) implies in addition that
the points~$m_\beta^2$ are absolute minima of the function~$a+b$. Moreover,
our theorem makes no statement on the condition~(ii). Nevertheless, the above
theorem shows that our Lorentz invariant variational principle is closely related
to state stability, and it seems a promising strategy for satisfying state stability
to construct minimizers of our variational principle, and then to select those
minimizers which satisfy the additional conditions of Definition~\ref{def611}.

In order to get stronger conditions from our variational principle,
one needs to consider more general variations of the fermionic projector.
A convenient method is to consider a ``test Dirac sea'' of infinitesimal weight
and of possibly negative mass. Thus we consider instead of~(\ref{hPdef}) the
fermionic projector
\[ \hat{P}(k) \;=\; \sum_{\beta=1}^{g+1} \rho_\beta\; (k\slsh+m_\beta)\, \delta(k^2-m_\beta^2)
\, \Theta(-k^0) \]
and assume that~$\rho_{g+1}$
vanishes for our unperturbed fermionic projector. We vary~$\rho_{g+1}$ for given~$m_{g+1}
\in \R$ and introduce the function~$V$ by
\[ V(m_{g+1}) \;=\; \frac{1}{2 \,m_{g+1}^3}\: \frac{\partial}{\partial \rho_{g+1}} {\mathcal{S}}
\big|_{\rho_{g+1}=0}\:. \]
We refer to~$V$ as the {\bf{variation density}}.
We point out that the~$(g+1)^{\mbox{\scriptsize{st}}}$ Dirac sea should not be considered
as being of physical significance; it is merely a mathematical tool to generate
variations of the action.
\begin{Thm} \label{thm2}
The functions~$a$ and~$b$ in~(\ref{62f}) are related to the variation density by
\[ V(m) \;=\; \epsilon(m)\, a(m^2)\,+\, b(m^2) \:. \]
\end{Thm}
{\Proof} Taking into account that~$m_{g+1}$ may be negative, (\ref{v2}) is modified to
\[ \frac{1}{2} \int_0^\infty \Tr \left( \hat{Q}(q)  \: \hat{P}(q) \right) b\: db 
\;=\; 2 \sum_{\beta=1}^g \rho_\beta\: m_\beta^3 \left(\epsilon(m_\beta)\, a(m_\beta^2)
\,+\, b(m_\beta^2) \right) . \]
Differentiating with respect to~$\rho_g$ gives the result.
\QED
In view of this theorem, the conditions~(ii) and~(iii) are can be expressed in terms
of the variation density by
\begin{description}
\item[(ii')] $V(m) \;\geq\; V(-m) \qquad\quad \forall m \in \R^+$
\item[(iii')] $V(m_\beta) \;\leq\; \inf_{\R^+} V \qquad\;\, \forall \beta \in \{1, \ldots g\}$.
\label{pageiii}
\end{description}

\section{Convolutions of Lorentz Invariant Distributions}
\setcounter{equation}{0}  \label{sec5}
We saw in the previous section that it is useful to rewrite products in position space
as convolutions in momentum space,
\beq \label{cint}
\widehat{F G}(q) \;=\; (\hat{F} * \hat{G}) := \int \frac{d^4k}{(2 \pi)^4}\:
\hat{F}(k)\, \hat{G}(q-k) \:,
\eeq
and to analyze the resulting convolution integrals (see~(\ref{hMdistr}) and~(\ref{hQdef})).
In this section we study such convolution integrals systematically.
The distributions of interest like~(\ref{Fhdef}) or~(\ref{hFkdef})
are Lorentz invariant and are supported in the mass cone~${\mathcal{C}}$.
It is helpful to decompose these distributions into the sum of a distribution supported in the upper
mass cone and a distribution supported in the lower mass cone.
This gives rise to the following definition.
\begin{Def}
A Lorentz-invariant distribution~$\hat{F} \in {\mathcal{S}}'(\hat{M})$ is called {\bf{negative}}
if it is supported in ${\mathcal{C}}^{\wedge}$ and can be written as
\beq \label{decomp}
\hat{F}(k) \;=\; f(k^2)\, \Theta(k^2)\: \Theta(-k^0)
\eeq
with a real-valued distribution~$f \in {\mathcal{S}}'(\R^+)$. Similarly,
$\hat{F}(k)$ is called {\bf{positive}} if~$\hat{F}(-k)$ is negative.
\end{Def}
We remark for clarity that the representation~(\ref{decomp}) poses an extra condition
which in particular excludes all distributions supported at the origin.
For example, the distribution~$\delta^4(k)$ is Lorentz invariant, supported
in~${\mathcal{C}}^\wedge$, but it cannot be represented in the form~(\ref{decomp}).

Since~$\hat{F}$ is completely characterized by~$f$, it is most convenient to work
only with the small letters.
Note that for notational convenience, the function~$f$ does not carry a hat;
to clarify momentum space we shall write out the arguments $a:=k^2, \ldots$.
In our computations we will use a compact notation which reminds of the corresponding
operations in position space. The following list shows what
operations on~$f$ and~$g$ mean for the corresponding distributions~$\hat{F}$ and~$\hat{G}$:
\beq \label{rules}
\left. \begin{array}{ccc}
(f \cdot g)(a) &\quad \Longleftrightarrow \quad& (F * G)(k) \\
\overline{f}(a) &\Longleftrightarrow& F(-k) \\
\Pdd f(a) &\Longleftrightarrow& i k\slsh \,F(k)
\end{array} \qquad \right\}
\eeq
The notation~$\overline{f}$ cannot lead to confusion because~$f(a)$ is real-valued,
and therefore the only reasonable meaning is to take the complex conjugate in position
space, which corresponds to flipping the sign of~$k$.
If~$f$ is negative, then clearly $\overline{f}$ is positive. Keeping this in mind, we
shall always assume that~$f$ and~$g$ correspond to negative distributions, whereas
positive distributions are denote by~$\overline{f}$ and~$\overline{g}$.
The operator~$\Pdd$ is convenient for handling Dirac matrices.

\begin{figure}[tb]
\begin{center}
\begin{picture}(0,0)%
\includegraphics{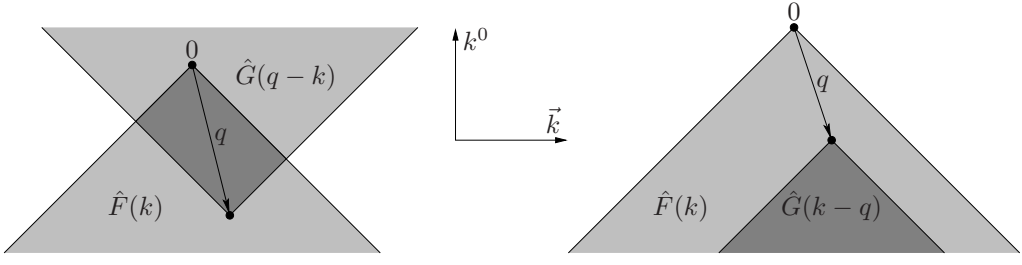}%
\end{picture}%
\setlength{\unitlength}{2072sp}%
\begingroup\makeatletter\ifx\SetFigFont\undefined%
\gdef\SetFigFont#1#2#3#4#5{%
  \reset@font\fontsize{#1}{#2pt}%
  \fontfamily{#3}\fontseries{#4}\fontshape{#5}%
  \selectfont}%
\fi\endgroup%
\begin{picture}(12174,3042)(-11,-1873)
\put(3376,164){\makebox(0,0)[b]{\smash{{\SetFigFont{10}{12.0}{\rmdefault}{\mddefault}{\updefault}{$\hat{G}(q-k)$}%
}}}}
\put(2521,-511){\makebox(0,0)[lb]{\smash{{\SetFigFont{10}{12.0}{\rmdefault}{\mddefault}{\updefault}{$q$}%
}}}}
\put(1576,-1411){\makebox(0,0)[b]{\smash{{\SetFigFont{10}{12.0}{\rmdefault}{\mddefault}{\updefault}{$\hat{F}(k)$}%
}}}}
\put(2251,479){\makebox(0,0)[b]{\smash{{\SetFigFont{10}{12.0}{\rmdefault}{\mddefault}{\updefault}{0}%
}}}}
\put(5491,569){\makebox(0,0)[lb]{\smash{{\SetFigFont{10}{12.0}{\rmdefault}{\mddefault}{\updefault}{$k^0$}%
}}}}
\put(6481,-421){\makebox(0,0)[lb]{\smash{{\SetFigFont{10}{12.0}{\rmdefault}{\mddefault}{\updefault}{$\vec{k}$}%
}}}}
\put(9901,-1411){\makebox(0,0)[b]{\smash{{\SetFigFont{10}{12.0}{\rmdefault}{\mddefault}{\updefault}{$\hat{G}(k-q)$}%
}}}}
\put(8101,-1411){\makebox(0,0)[b]{\smash{{\SetFigFont{10}{12.0}{\rmdefault}{\mddefault}{\updefault}{$\hat{F}(k)$}%
}}}}
\put(9721,119){\makebox(0,0)[lb]{\smash{{\SetFigFont{10}{12.0}{\rmdefault}{\mddefault}{\updefault}{$q$}%
}}}}
\put(9451,929){\makebox(0,0)[b]{\smash{{\SetFigFont{10}{12.0}{\rmdefault}{\mddefault}{\updefault}{0}%
}}}}
\end{picture}%
\caption{Convolution of two negative distributions (left)
and a mixed convolution (right).}
\label{figposneg}
\end{center}
\end{figure}

If~$\hat{F}$ and~$\hat{G}$ are both negative (or both positive), the integrand of
the convolution integral~(\ref{cint}) is compact (see Figure~\ref{figposneg} (left)).
If on the other hand one factor is positive and the other is negative, the support of
the integrand will in general be non-compact (see Figure~\ref{figposneg} (right)).
These two cases are essentially different and will be treated separately
in Subsections~\ref{subneg} and~\ref{submixed}.
In Subsection~\ref{sechM} we shall analyze the distribution~$\hM$,
which already appeared in~(\ref{Mtdef}, \ref{tMsymm}).

\subsection{Convolutions of Negative Distributions} \label{subneg}
We introduce the function $\Delta(a,b,c)$ by
\begin{equation} \label{defDelta}
\Delta \;=\; a^2 + b^2 + c^2 - 2 (ab + ac + bc) \:.
\end{equation}
Note that $\Delta$ is symmetric in its three arguments.
\begin{Lemma} \label{lemmapos}
Suppose that $f$ and $g$ are negative distributions.
Then the following convolutions are also negative and are given explicitly by
\begin{eqnarray}
(f \cdot g)(a) &=& \frac{1}{32 \pi^3} \int_0^a dc \:f(c) \int_0^{(\sqrt{a}
-\sqrt{c})^2} db \:g(b)\; \frac{\sqrt{\Delta}}{a} \label{cp1} \\
(\Pdd f \cdot g)(a) &=& (\Pdd \alpha)(a) \;, \nonumber \\
\alpha(a) &=& \frac{1}{32 \pi^3} \int_0^a dc \:f(c) \int_0^{(\sqrt{a}
-\sqrt{c})^2} db \:g(b)\:\sqrt{\Delta}\; \frac{a-b+c}{2 a^2} \label{cp2} \\
(f \cdot \Pdd g)(a) &=& (\Pdd \beta)(a) \;, \nonumber \\
\beta(a) &=& \frac{1}{32 \pi^3} \int_0^a dc \:f(c) \int_0^{(\sqrt{a}
-\sqrt{c})^2} db \:g(b)\:\sqrt{\Delta}\; \frac{a+b-c}{2 a^2} \label{cp3} \\
(\partial_k f \cdot \partial^k g)(a) &=&
\frac{1}{32 \pi^3} \int_0^a dc \:f(c) \int_0^{(\sqrt{a}
-\sqrt{c})^2} db \:g(b)\:\sqrt{\Delta}\; \frac{c+b-a}{2 a} \:.
\label{cp4}
\end{eqnarray}
\end{Lemma}
{\Proof} Clearly, the convolutions are again Lorentz invariant, and a geometric argument (see Figure~\ref{figposneg}~(left)) immediately yields that they are again negative.
To calculate the convolutions, we choose any $q \in {\mathcal{C}}^\wedge$ with $q^2=a$. Then
\begin{eqnarray*}
\lefteqn{ (f \cdot g)(a) \;=\; \int \frac{d^4k}{(2 \pi)^4}\:
\hat{F}(k)\: \hat{G}(q-k) } \\
&=& \int_0^\infty dc \int_0^\infty db \: f(c)\: g(b)\:
\int \frac{d^4 k}{(2 \pi)^4}\:
\delta(k^2-c)\:\Theta(-k^0)\; \delta((q-k)^2-b)\: \Theta(k^0-q^0) \\
&=& \int_0^\infty dc \int_0^\infty db \: f(c)\: g(b)\:
\int \frac{d^4k}{(2 \pi)^4}\: \delta(k^2-c)\:\Theta(-k^0)\;
\delta(q^2 -2qk+ c-b)\: \Theta(k^0-q^0) \:.
\end{eqnarray*}
Due to Lorentz symmetry, we can assume that $q$ points in the
direction of $k^0$. 
Introducing polar coordinates $(\omega=k^0, p=|\vec{k}|, \Omega \in S^2)$,
we can carry out the momentum integrals,
\begin{eqnarray}
(f \cdot g)(a) &=& \frac{1}{4 \pi^3}
\int_0^\infty dc \int_0^\infty db \: f(c)\: g(b) \nonumber \\
&&\quad \times \int_{q_0}^0 d\omega \int_0^\infty p^2\: dp \;
\delta(\omega^2 - p^2 -c)\:
\delta \!\left(a +2 \omega \sqrt{a} + c-b
\right) \nonumber \\
&=& \frac{1}{4 \pi^3} \int_0^\infty dc \int_0^\infty db \:
f(c)\: g(b) \;\Theta(\omega_0^2-c)\: \chi_{[-\sqrt{a},0]}(\omega_0)\:
\frac{p_0}{4 \sqrt{a}} \:, \label{25a}
\end{eqnarray}
where~$\chi$ is the characteristic function, and~$\omega_0$ and~$p_0$ are determined
by the conditions
\begin{equation} \label{op0}
\omega_0 \;=\; \frac{c+a-b}{-2 \sqrt{a}}\;,\qquad
p_0^2 \;=\; \omega_0^2 - c\:.
\end{equation}
A short calculation shows that~$p_0^2=\Delta/(4 a)$ with
$\Delta$ according to~(\ref{defDelta}). Using that the Heaviside and the characteristic
function in~(\ref{25a}) vanish unless $\sqrt{a} \geq \sqrt{c}+\sqrt{b}$,
we obtain~(\ref{cp1}).

In order to derive~(\ref{cp2}), we first compute
\begin{eqnarray*}
(\Pdd f \cdot g)(a) &=& \int \frac{d^4k}{(2 \pi)^4}\:
i k\slsh \,\hat{F}(k)\: \hat{G}(q-k) \\
&=& \int_0^\infty dc \int_0^\infty db \: f(c)\: g(b)\:
\int \frac{d^4 k}{(2 \pi)^4}\:i k \slsh\:
\delta(k^2-c)\: \delta(q^2 -2qk+ c-b)\:.
\end{eqnarray*}
We again choose a reference frame with $q=(-\sqrt{a}, \vec{0})$ and
introduce polar coordinates~$(\omega=k^0, p=|\vec{k}|,
\Omega \in S^2)$. Then
\[ (\Pdd f \cdot g)(a) \;=\;
\frac{1}{4 \pi^3} \int_0^\infty dc \int_0^\infty db \:
f(c)\: g(b) \;i \omega_0 \gamma^0\: \Theta(\omega_0^2-c)\: \1_{[-\sqrt{a},0]}(\omega_0)\:
\frac{p_0}{4 \sqrt{a}} \]
with $\omega_0$ and $p_0$ according to~(\ref{op0}). Note that, due to spherical
symmetry, the contribution $\sim \vec{\gamma}$ vanishes.
A straightforward computation yields
\[ (\Pdd f \cdot g)(a) \;=\;
\frac{1}{32 \pi^3} \int_0^a dc \int_0^{(\sqrt{a}-\sqrt{c})^2} db \:
f(c)\: g(b) \;\left(-i \sqrt{a}\:\gamma^0\right)\:
\frac{a-b+c}{2a}\: \frac{\sqrt{\Delta}}{a} \;=\;
(\Pdd \alpha)(a) \]
with $\alpha$ as in~(\ref{cp2}). The relation~(\ref{cp3})
immediately follows from~(\ref{cp2}) using the symmetry in $f$ and $g$.

To derive~(\ref{cp4}), we again assume that $q$ points into the direction of
$k^0$. Then, exploiting the spherical symmetry,
\begin{eqnarray*}
\lefteqn{ (\partial_k f \cdot \partial^k g)(a)
\;=\; \int \frac{d^4k}{(2 \pi)^4}\:
i k_l \,\hat{F}(k)\: i (q^l - k^l) \:\hat{G}(q-k) } \\
&=& \int_0^\infty dc \int_0^\infty db \: f(c)\: g(b)\:
\int \frac{d^4 k}{(2 \pi)^4}\: (-k q + c)\:
\delta(k^2-a)\: \delta(q^2 -2qk+ c-b) \\
&=& \frac{1}{4 \pi^3} \int_0^\infty dc \int_0^\infty db \:
f(c)\: g(b) \;(\omega_0\:\sqrt{a} + c)\: \Theta(\omega_0^2-c)\: \chi_{[-\sqrt{a},0]}(\omega_0)\:
\frac{p_0}{4 \sqrt{a}}\:,
\end{eqnarray*}
and a short computation gives~(\ref{cp4}).
\QED

\subsection{Mixed Convolutions}  \label{submixed}
We shall now analyze the convolution $f \cdot \overline{g}$ of a positive and
a negative distribution.
We begin with the case $q \in {\mathcal{C}}^\wedge$ where the momentum is in the lower
mass cone. This case is unproblematic because the integrand of the convolution integral~(\ref{cint})
becomes compact if we assume that~$f(k^2)$ vanishes identically for sufficiently large~$k^2$.
Making this additional assumption, we can use the same method as in Lemma~\ref{lemmapos}.

\begin{Lemma} \label{lemmamix}
Suppose that $f$ and $g$ are negative distributions and that~$f(k^2)$ 
vanishes identically for large~$k^2$.
Then for $q \in {\mathcal{C}}^\land$ and setting $a=q^2 \geq 0$, the following convolutions are
well-defined and given explicitly by
\begin{eqnarray}
(f \cdot \overline{g})(a) &=& \frac{1}{32 \pi^3}
\int_a^\infty dc\: f(c) \int_0^{(\sqrt{c}-\sqrt{a})^2} db \: g(b) \;
\frac{\sqrt{\Delta}}{a} \label{cp5} \\
(\Pdd f \cdot \overline{g})(a) &=& (\Pdd \alpha)(a) \;, \nonumber \\
\alpha(a) &=& \frac{1}{32 \pi^3} \int_a^\infty dc\: f(c) \int_0^{(\sqrt{c}-\sqrt{a})^2}
db \:g(b)\:\sqrt{\Delta}\; \frac{a-b+c}{2 a^2} \label{cm2} \\
(f \cdot \Pdd \overline{g})(a) &=& (\Pdd \beta)(a) \;, \nonumber \\
\beta(a) &=& \frac{1}{32 \pi^3} \int_a^\infty dc\: f(c) \int_0^{(\sqrt{c}-\sqrt{a})^2}
db \:g(b)\:\sqrt{\Delta}\; \frac{a+b-c}{2 a^2} \label{cm3} \\
(\partial_k f \cdot \partial^k \overline{g})(a) &=&
\frac{1}{32 \pi^3}
\int_a^\infty dc  \: f(c)\int_0^{(\sqrt{c}-\sqrt{a})^2} db\: g(b) \; \sqrt{\Delta}\; \frac{c+b-a}{2 a} \:. \label{cp8}
\end{eqnarray}
\end{Lemma}
{\Proof} The momentum vector $q \in {\mathcal{C}}^\land$ satisfies the
inequality $q^0<0$ and thus
\begin{eqnarray}
\lefteqn{ (f \cdot \overline{g})(a) \;=\; \int \frac{d^4k}{(2 \pi)^4}\:
\hat{F}(k)\: \hat{G}(k-q) } \label{convform} \\
&=& \int_0^\infty dc \int_0^\infty db \: f(c)\:g(b)\:
\int \frac{d^4 k}{(2 \pi)^4}\:
\delta(k^2-c)\:\Theta(-k^0)\; \delta((k-q)^2-b)\: \Theta(q^0-k^0) \nonumber \\
&=& \int_0^\infty dc \int_0^\infty db \: f(c)\: g(b)\:
\int \frac{d^4k}{(2 \pi)^4}\: \delta(k^2-c)\;
\delta(c-2qk+q^2-b)\: \Theta(q^0-k^0) \:. \nonumber
\end{eqnarray}
Due to Lorentz invariance, we can assume that $q=(-\sqrt{a}, \vec{0})$.
Again choosing polar coordinates~$(\omega, p, \Omega)$, we obtain
\begin{eqnarray*}
(f \cdot \overline{g})(a) &=& \frac{1}{4 \pi^3}
\int_0^\infty dc \int_0^\infty db \: f(c)\: g(b) \\
&&\quad \times \int_{-\infty}^{-\sqrt{a}} d\omega
\int_0^\infty p^2\, dp \;\delta(\omega^2 - p^2 - c)\:
\delta \!\left(c +2 \omega \sqrt{a} + a-b \right) \\ 
&=& \frac{1}{8 \pi^3}
\int_0^\infty dc \int_0^\infty db \: f(c)\: g(b) \\
&&\quad \times \int_{-\infty}^{-\sqrt{a}} d\omega\;
\Theta(\omega^2-c)\: \sqrt{\omega^2-c} \;\delta(\omega^2 - p^2 - c)\:
\delta \!\left(c +2 \omega \sqrt{a} + a-b \right) \\
&=& \frac{1}{8 \pi^3}
\int_0^\infty dc \int_0^\infty db \: f(c)\: g(b) \;
\sqrt{\omega_0^2-c}\: \frac{1}{2 \sqrt{a}}\:
\Theta(\omega_0^2-c)\: \Theta(c-b-a) \:,
\end{eqnarray*}
where we set
\[ \omega_0 \;=\; \frac{a-b+c}{-2 \sqrt{a}}\:. \]
A straightforward calculation gives~(\ref{cp5}).

The relations~(\ref{cm2})--(\ref{cp8}) are obtained with the same
method as the corresponding relations~(\ref{cp2})--(\ref{cp4})
in Lemma~\ref{lemmapos}.
\QED
This lemma can also be applied in the case when~$q$ in the upper mass cone~${\mathcal{C}}^\vee$.
Namely, changing the variables~$k-q \rightarrow q$ in~(\ref{convform}), one sees that
flipping the sign of~$q$ corresponds to exchanging~$f$ and~$g$.
Hence the corresponding formulas are obtained if in~(\ref{cp5}, \ref{cp8}) we
exchange~$f$ and~$g$ on the right.

We come to the most interesting case that $q$ lies outside the mass cone, $q^2<0$.
Since the hyperbolas~$\{k^2=c,\: k^0<0\}$ and~$\{(k-q)^2=c',\: (k^0-q^0)<0\}$
(for $c, c' > 0$) have a non-compact intersection, in this case the convolution
integral~(\ref{cint}) necessarily diverges for any non-trivial
distributions~$\hat{F}$ and~$\hat{G}$.
Our method to bypass this problem is to first break the Lorentz invariance
by a regularization. This makes the convolution integral~(\ref{cint}) finite.
Then we will subtract suitable counter terms which are supported on the light cone.
After that, we can remove the regularization. We thus obtain Lorentz invariant formulas
for the convolution integrals, which have to be understood ``modulo infinite contributions
supported on the light cone.'' This method of disregarding contributions on
the light cone is motivated from the fact that we computed
the pointwise product in~(\ref{Adef}) only away from the light cone,
with the intention of then extending the resulting function in the distributional sense across the light
cone~(\ref{Mtdef}). The resulting distributions depend only on~(\ref{Adef})
away from the light cone. Therefore, it is irrelevant to us how the regularized quantities
behave on the light cone as the regularization is removed. This is why we can simply
subtract divergent terms which are supported on the light cone.

To regularize~$\hat{F}$, we set
\beq
(f^\varepsilon)(k) \;=\; f(k^2)\: e^{\varepsilon k^0} \label{r1}
\eeq
and introduce the distribution~$\hat{F} \in {\mathcal{S}}'(\hat{M})$ in analogy
to~(\ref{decomp}) by
\beq \label{r2}
\hat{F}^\varepsilon(k) \;=\; f^\varepsilon(k)\, \Theta(k^2)\: \Theta(-k^0) \:.
\eeq
In this way, we can use the previous notation, with the only difference that
the functions~$f^\varepsilon$ and~$g^\varepsilon$ are no longer Lorentz invariant
and are thus no longer functions of~$q^2$, but of~$q \in \hat{M}$.
For simplicity, we assume that~$f$ and~$g$ have compact support.
\begin{Lemma} \label{lemma32}
Suppose that $f(k^2)$ and $g(k^2)$ are negative distributions which vanish identically
for large~$k^2$.
Then for~$q \in \hat{M} \setminus {\mathcal{C}}$ and setting~$a=q^2 \leq 0$,
the following formulas hold for the products
of the corresponding regularized distributions~(\ref{r1}, \ref{r2}),
\begin{eqnarray}
(f^\varepsilon \cdot \overline{g^\varepsilon})(q)
&=& \frac{1}{32 \pi^3} \int_0^\infty dc\: f(c)
\int_0^\infty db\: g(b)\: H_\varepsilon(q,b,c) \label{E1rel} \\
(\partial_k f^\varepsilon \cdot \partial^k \overline{g^\varepsilon})(q)
&=& \frac{1}{32 \pi^3} \int_0^\infty dc\: f(c)
\int_0^\infty db\: g(b)\;
\frac{b+c-a}{2}\: H_\varepsilon(q,b,c) \:, \label{E3rel}
\end{eqnarray}
where the function $H_\varepsilon$ is given by
\beq \label{E2rel}
H_\varepsilon(q,b,c) \;=\; \frac{1}{2 \varepsilon|\vec{q}|}\;
\exp \left(\varepsilon|\vec{q}|\: \frac{\sqrt{\Delta}}{a}
+ \varepsilon q^0\: \frac{c-b}{a} \right) .
\eeq
\end{Lemma}
{\Proof} Introducing the vector $u=(\varepsilon, \vec{0})$, we can write
the product $f^\varepsilon \overline{g^\varepsilon}$ as
\[ (f^\varepsilon \cdot \overline{g^\varepsilon})(q)
\;=\; \int_0^\infty dc\: f(c) \int_0^\infty db\: g(b)\: I(q,b,c) \]
with
\[ I(q,b,c) \;=\; \int \frac{d^4 k}{(2 \pi)^4}\: \delta(k^2-c)\:
\delta((k-q)^2-b)\: \Theta(-k^0)\:\Theta(q^0-k^0)\: e^{ku + (k-q) u} \:. \]
We choose a reference frame where $q=(0,x,0,0)$ and
$u=(\alpha, \beta, 0, 0)$ with $x>0$ and $\alpha > |\beta|$. Choosing
cylindrical coordinates~$(\omega, p, r, \varphi)$ with
$k^0=\omega$ and $\vec{k} = (p, r \cos \varphi, r \sin \varphi)$,
we can compute the integrals,
\begin{eqnarray*}
I(q,b,c) &=& \frac{1}{8 \pi^3}
\int_{-\infty}^0 d\omega \int_{-\infty}^\infty dp \int_0^\infty r\:dr \\
&& \times\: \delta(\omega^2-p^2-r^2-c)\: \delta(\omega^2 - (p-x)^2 - r^2 - b)\: e^{2 \alpha \omega - \beta (2 p-x)} \\
&=& \frac{1}{16 \pi^3} \int_{-\infty}^0 d\omega
\int_{-\infty}^\infty dp\: \Theta(\omega^2-p^2-c)\: \delta(2px-x^2+c-b)\:
e^{2 \alpha \omega - \beta (2 p-x)} \\
&=& \frac{1}{32 \pi^3}\: \frac{1}{\alpha}
\int_{-\infty}^\infty dp\: \delta(2px-x^2+c-b)\: e^{-2 \alpha \sqrt{p^2+c} - \beta (2p-x)} \\
&=& \frac{1}{64 \pi^3}\:\frac{1}{\alpha x} \: e^{-2 \alpha \sqrt{K^2+c}
- \beta (2K-x)}\: ,
\end{eqnarray*}
where we set
\[ K \;=\; \frac{x^2-c+b}{2x} \:. \]
This can also be written as
\begin{equation} \label{Iform}
I(q,b,c) \;=\; \frac{1}{32 \pi^3}\: \frac{1}{2 \alpha x}\:
\exp \left(- \alpha x\:A - \beta x\: B \right) ,
\end{equation}
where $A$ and $B$ are the expressions
\begin{eqnarray*}
A &=& 2 \:\frac{\sqrt{K^2+c}}{x} \;=\; -\frac{\sqrt{(-a-c+b)^2 - 4ac}}{a}
\;=\; -\frac{\sqrt{\Delta}}{a} \\
B &=& \frac{2 K - x}{x} \;=\; \frac{c-b}{a} \:,
\end{eqnarray*}
which are covariant and independent of the regularization.
Next, we transform back to the original reference frame where
$q=(q^0, \vec{q})$ and $u=(\varepsilon, \vec{0})$. Since
$\beta x = -u q = -\varepsilon q^0$ and
$\alpha x = \sqrt{-u^2 q^2 + (u q)^2} = \varepsilon |\vec{q}|$,
we have the simple transformation rules
\beq \label{trafo1}
\alpha x \to \varepsilon |\vec{q}| \;,\spc
\beta x \to -\varepsilon q^0\:.
\eeq
Applying these transformation rules in~(\ref{Iform}) gives~(\ref{E2rel}).

The relation~(\ref{E3rel}) follows immediately from the transformations
\begin{eqnarray*}
(\partial_k f^\varepsilon \cdot \partial^k
\overline{g^\varepsilon})(q) &=&
\int_0^\infty dc \int_0^\infty db\:f(c) \:g(b)\: K(q,b,c)
\qquad {\mbox{with}} \\
K(q,b,c) &=& \int \frac{d^4k}{(2 \pi)^4}\: k_l\: (k-q)^l\:
\delta(k^2-c)\: \delta((k-q)^2-b)\: \Theta(k^0)\: e^{ku + (k-q) u} \\
&=& -\int \frac{d^4k}{(2 \pi)^4}\: \frac{1}{2}\left(
(k-(k-q))^2 - k^2 - (k-q)^2 \right) \\
&&\hspace*{2cm} \times
\delta(k^2-c)\: \delta((k-q)^2-b)\: \Theta(k^0)\: e^{ku + (k-q) u} \\
&=& -\frac{1}{2}\:(a-c-b)\: I(q,b,c)\:.
\end{eqnarray*}

\vspace*{-1.8em}
\QED

Obviously, the function $H_\varepsilon$, (\ref{E2rel}), becomes singular
as $\varepsilon \searrow 0$. The next lemma shows that after dropping
suitable terms which are supported on the light cone, the limit
$\varepsilon \searrow 0$ exists and is Lorentz invariant.

\begin{Lemma} \label{lemma33}
Suppose that $f(k^2)$ and $g(k^2)$ are negative distributions which vanish identically
for large~$k^2$.
Then the products of the corresponding regularized distributions~(\ref{r1})
have the decomposition
\begin{eqnarray*}
(f^\varepsilon \cdot \overline{g^\varepsilon})(q) &=& S_1^\varepsilon(q)
+ R_1^\varepsilon(q) \\
(\partial_k f^\varepsilon \cdot \partial^k \overline{g^\varepsilon})(q) &=& S_2^\varepsilon(q) + R_2^\varepsilon(q) \:,
\end{eqnarray*}
where the $S^\varepsilon_i$ are distributions which are supported
on the light cone,
\[ {\mbox{\rm{supp}}}\, S^\varepsilon_i \:\subset\: \{ \xi^2=0 \} \:, \]
and the $R^\varepsilon_i$ are regular as $\varepsilon \searrow 0$. The limits
\[ R_i \;=\; \lim_{\varepsilon \searrow 0} R^\varepsilon_i \]
are the Lorentz invariant distributions
\beq \label{rjdef}
R_i(q) \;=\; \int_0^\infty dc\: f(c) \int_0^\infty db\: g(b)\: K_i(q,b,c)
\eeq
with
\begin{eqnarray}
K_1(q,b,c) &=& \frac{1}{32 \pi^3} \left\{
\begin{array}{cl} \displaystyle \frac{\sqrt{\Delta}}{a}\: \Theta(\sqrt{b} -
\sqrt{a} - \sqrt{c}) - \frac{|b-c|}{a}\: \Theta(b-c) &
{\mbox{if $q \in {\mathcal{C}}^\lor$}} \\[.8em]
\displaystyle \frac{\sqrt{\Delta}}{a}\: \Theta(\sqrt{c} -
\sqrt{a} - \sqrt{b}) - \frac{|b-c|}{a}\: \Theta(c-b) &
{\mbox{if $q \in {\mathcal{C}}^\land$}} \\[.8em]
\displaystyle 
\frac{\sqrt{\Delta} - |b-c|}{2a} &
{\mbox{if $q \not \in {\mathcal{C}}$}} \end{array} \right.\qquad \label{K1def} \\
K_2(q,b,c) &=&
\frac{1}{32 \pi^3} \nonumber \\
&&\hspace*{-2.5cm} \times \left\{
\begin{array}{cl} \displaystyle \frac{\sqrt{\Delta}\: (b+c-a)}{2a}\: \Theta(\sqrt{b} -
\sqrt{a} - \sqrt{c}) - \frac{|b-c|\: (b+c)}{2a}\: \Theta(b-c) &
{\mbox{if $q \in {\mathcal{C}}^\lor$}} \\[.8em]
\displaystyle \frac{\sqrt{\Delta}\: (b+c-a)}{2a}\: \Theta(\sqrt{c} -
\sqrt{a} - \sqrt{b}) - \frac{|b-c|\:(b+c)}{2a}\: \Theta(c-b) &
{\mbox{if $q \in {\mathcal{C}}^\land$}} \\[.8em]
\displaystyle 
\frac{\sqrt{\Delta}\:(b+c-a) - |b-c|\:(b+c)}{4a} &
{\mbox{if $q \not \in {\mathcal{C}}\:,$}} \end{array} \right.\qquad \label{K2def}
\end{eqnarray}
where we again set~$a = q^2$.
\end{Lemma}
It is interesting to compare the result of this lemma in the case~$q \in {\mathcal{C}}^\wedge$
with Lemma~\ref{lemmamix}. Then the first summands in~(\ref{K1def}) and~(\ref{K2def})
coincide precisely with~(\ref{cp5}) and~(\ref{cp8}), respectively. The appearance of
the additional second summands in~(\ref{K1def}) and~(\ref{K2def}) is somewhat surprising;
it shows that extending the convolution integral in the distributional sense to~$q \in \hat{M}$ yields
additional contributions inside the mass cone. These additional contributions can be
understood from the fact they remove the poles of the convolution integral
on the cone~$\{q \:|\: q^2=0\}$. Namely, using the expansion
\[ \sqrt{\Delta} \;=\; |b-c| + {\cal{O}}(a)\:, \]
one sees that~$K_1$ and~$K_2$ have no poles at~$a=0$. \\[.5em]
{\em{Proof of Lemma~\ref{lemma33}.}} We first consider the product $f^\varepsilon \cdot \overline{g^\varepsilon}$. Extending the function
$H_\varepsilon$, (\ref{E2rel}), by zero to the interior mass cone $q \in {\mathcal{C}}$,
we obtain the expansion
\beq \label{Fex}
H_\varepsilon(q,b,c) \;=\; \frac{\Theta(-a)}{2 \varepsilon |\vec{q}|}
+\frac{q^0}{2 |\vec{q}|}\: \frac{\Theta(-a)}{a}\: (c-b)
+ \frac{\sqrt{\Delta}}{2a}\: \Theta(-a) + {\cal{O}}(\varepsilon)\:.
\eeq
Here the poles on the mass cone have a mathematical meaning as
distributional derivatives of the regular distribution~$\Theta(-a) \log a$. We implicitly use
this procedure throughout the proof.
The first two summands in~(\ref{Fex}) are not Lorentz invariant, and we
must analyze their behavior in position space. In preparation, we
derive a useful formula for spherically symmetric distributions which are supported on the light cone. In position space, such a distribution can
clearly be written in the form $g(\xi^0)\: \delta(\xi^2)$. We compute
its Fourier transform in polar coordinates $x=(t, r, \vartheta, \varphi)$,
\begin{eqnarray*}
f(q) &:=& \int d^4 \xi\: g(\xi^0) \:\delta(\xi^2)\: e^{-iqx} \\
&=& 2 \pi \int_{-\infty}^\infty dt \: g(t)\: e^{-i q^0 t}
\:\int_0^\infty r^2\: dr\: \delta(t^2-r^2)
\int_{-1}^1 d\cos \vartheta\: e^{i |\vec{q}| r \cos \vartheta} \\
&=& \frac{2 \pi}{i |\vec{q}|} \int_{-\infty}^\infty dt \: g(t)\:
e^{-i q^0 t} \:\int_0^\infty r\: dr\: \delta(t^2-r^2)
\left( e^{i |\vec{q}| r} - e^{-i |\vec{q}| r} \right) \\
&=& \frac{\pi}{i |\vec{q}|} \int_{-\infty}^\infty dt \: g(t)\:
e^{-i q^0 t} \left( e^{i |\vec{q}| |t|} - e^{-i |\vec{q}| |t|} \right) \\
&=& \frac{\pi}{i |\vec{q}|} \int_{-\infty}^\infty dt \: g(t)\:\epsilon(t)\:
e^{-i q^0 t} \left( e^{i |\vec{q}| t} - e^{-i |\vec{q}| t} \right)
\;=\; \frac{1}{2|\vec{q}|} \left( h(q^0+|\vec{q}|) - h(q^0 - |\vec{q}|) \right),
\end{eqnarray*}
where $h$ is the Fourier transform of the function
$2i \pi g(t) \epsilon(t)$. Since the function $g$ can be arbitrary,
the function $h$ is also arbitrary. Also, the above calculation clearly
carries through if $h$ is a one-dimensional distribution.
We conclude that for any distribution $h$, the expression
\begin{equation} \label{harmonic}
f(q) \;=\; \frac{1}{|\vec{q}|} \left( h(q^0+|\vec{q}|) -
h(q^0-|\vec{q}|) \right)
\end{equation}
is supported on the light cone. Choosing $h(x) = \epsilon(x)$, one sees
that the distribution
$2 \Theta(-a)/|\vec{q}|$, and thus also the first summand in~(\ref{Fex}),
are supported on the light cone. 
The second summand is not of the required form.
But choosing $h(x)=(c-b)\:\epsilon(x)/(4x)$, one sees that its extension
\[ f(q) \;=\; \frac{c-b}{2a} \:\times\: \left\{ \begin{array}{cl}
\displaystyle \frac{q^0}{|\vec{q}|} &
{\mbox{if $q \not \in {\mathcal{C}}$}} \\[0.8em]
-1 & {\mbox{if $q \in {\mathcal{C}}^\lor$}} \\
1 & {\mbox{if $q \in {\mathcal{C}}^\land$}}
\end{array} \right. \]
is supported on the light cone. Setting
\[ S_\varepsilon \;=\; \frac{\Theta(-a)}{2 \varepsilon |\vec{q}|}
+ f(q) + \frac{|b-c|}{2}\: \frac{\mbox{PP}}{a} \:, \]
$S_\varepsilon$ is supported on the light cone, and the difference
$H_\varepsilon - S_\varepsilon$ has a Lorentz invariant limit,
\begin{eqnarray*}
\lim_{\varepsilon \searrow 0} (H_\varepsilon - S_\varepsilon) \;=\;
\left\{ \begin{array}{cl}
\displaystyle -\frac{|b-c|}{a} \:\Theta(b-c) & {\mbox{if $q \in {\mathcal{C}}^\lor$}}
\\[.8em]
\displaystyle -\frac{|b-c|}{a} \:\Theta(c-b) & {\mbox{if $q \in {\mathcal{C}}^\land$}}
\\[.8em]
\displaystyle \frac{\sqrt{\Delta}-|b-c|}{2a} & {\mbox{if $q \not \in {\mathcal{C}}$}}\:.
\end{array} \right.
\end{eqnarray*}
If $q \in {\mathcal{C}}^\land$, we must also take into account the contribution
from Lemma~\ref{lemmamix}. In the case $q \in {\mathcal{C}}^\lor$, we can apply
Lemma~\ref{lemmamix} after double conjugation,
\[ (f \cdot \overline{g})(q) \;=\; \overline{(g \cdot \overline{f})(-q)}
\;=\; \frac{1}{32 \pi^3}
\int_a^\infty dc\: g(c) \int_0^{(\sqrt{c}-\sqrt{a})^2} db \: f(b)
\; \frac{\sqrt{\Delta}}{a} \:. \]
Adding all these contributions gives the result for
the product $f^\varepsilon \cdot \overline{g^\varepsilon}$.

In order to treat the product $\partial_k f^\varepsilon \cdot
\partial^k \overline{g^\varepsilon}$, we note that~(\ref{cp8})
and~(\ref{E3rel}) differ from the corresponding formulas~(\ref{cp5})
and~(\ref{E1rel}) only by the additional factor $(b+c-a)/2$.
Since $b$ and $c$ were treated as fixed parameters throughout the proof,
this additional factor can be written as the differential operator
in position space $(b+c+\OBox_\xi)/2$. Using that the distributions
$\OBox_\xi S^\varepsilon_i$ are again supported on the light cone, the
above arguments all go through after applying the operator
$(b+c+\OBox_\xi)/2$. From this consideration we obtain~(\ref{rjdef}) with
\[ K_2(q,b,c) \;=\; K_1(q,b,c) \: \frac{b+c-a}{2}\:. \]
Finally, we subtract the contribution
\beq \label{conlc}
\left\{ \begin{array}{cl} \displaystyle \frac{|b-c|}{2}\: \Theta(b-c) &
{\mbox{if $q \in {\mathcal{C}}^\lor$}} \\[.8em]
\displaystyle \frac{|b-c|}{2}\: \Theta(c-b) &
{\mbox{if $q \in {\mathcal{C}}^\land$}} \\[.8em]
\displaystyle
\frac{|b-c|}{4} & {\mbox{if $q \not \in {\mathcal{C}}\:,$}} \end{array} \right.
\eeq
which for any fixed $b$ and $c$ can be written as
$c_1 + c_2 \: \Theta(q^2)\: \epsilon(q^0)$ with constants $c_1$ and~$c_2$.
The Fourier transform of the first summand is a multiple of the distribution~$\delta^4(\xi)$.
Using that~$\Box^2_q (\Theta(q^2)\: \epsilon(q^0))=0$ in the distributional sense,
one sees that the second summand is supported on the light cone
(alternatively, see Corollary~\ref{lemmaA1} below). We conclude that
the contribution~(\ref{conlc}) is supported on the light cone
$\{\xi \:|\: \xi^2=0\}$, completing the proof.
\QED
This lemma shows that the products $f \cdot \overline{g}$ and
$\partial_k f \cdot \partial^k \overline{g}$ are singular only on the
light cone. Furthermore, it gives a useful explicit formula for the regular
contribution away from the light cone in momentum space. Note that the
regularization is needed in order to make the singular contribution on
the light cone finite, but the remaining regular contribution does not depend
on the regularization. With this in mind, we write the result of Lemma~\ref{lemma33} in the compact form
\begin{eqnarray}
(f \cdot \overline{g})(q) &=& {\mbox{(l.c.)}} +
\int_0^\infty dc\: f(c) \int_0^\infty db\: g(b)\: K_1(q,b,c)
\label{rs1} \\
(\partial_k f \cdot \partial^k \overline{g})(q) &=& {\mbox{(l.c.)}} +
\int_0^\infty dc\: f(c) \int_0^\infty db\: g(b)\: K_2(q,b,c)\:,
\label{rs2}
\end{eqnarray}
where ``(l.c.)'' stands for a singular (possibly infinite)
contribution on the light cone.

The next lemma treats the products which involve one derivative.
\begin{Lemma} \label{lemma34}
Suppose that $f$ and $g$ are negative, Lorentz invariant distributions.
Then, using the short notation introduced before~(\ref{rs1}),
\begin{eqnarray}
(\Pdd f \cdot \overline{g})(q) &=& {\mbox{\rm{(l.c.)}}} +
\int_0^\infty dc\: f(c) \int_0^\infty db\: g(b)\: L_1(q,b,c)
\label{s1s} \\
(f \cdot \Pdd \overline{g})(q) &=& {\mbox{\rm{(l.c.)}}} +
\int_0^\infty dc\: f(c) \int_0^\infty db\: g(b)\: L_2(q,b,c)
\label{s2s}
\end{eqnarray}
with
\begin{eqnarray}
\lefteqn{ L_1(q,b,c) \;=\; \frac{i \qslsh}{32 \pi^3} } \nonumber \\
&&\times \left\{ \begin{array}{cl}
\displaystyle \sqrt{\Delta} \:\frac{a-b+c}{2a^2}\:
\Theta(\sqrt{b} - \sqrt{a} - \sqrt{c}) +
\frac{(b-c)^2-2ab}{2a^2}\: \Theta(b-c)
& {\mbox{if $q \in {\mathcal{C}}^\lor$}} \\[.8em]
\displaystyle\sqrt{\Delta} \:\frac{a-b+c}{2a^2}\:
\Theta(\sqrt{c} - \sqrt{a} - \sqrt{b}) -
\frac{(b-c)^2-2ab}{2a^2}\: \Theta(c-b)
& {\mbox{if $q \in {\mathcal{C}}^\land$}}\\[.8em]
\displaystyle \sqrt{\Delta} \:\frac{a-b+c}{4a^2}
+ \frac{(b-c)^2-2ab}{4a^2}\: \epsilon(b-c)
& {\mbox{if $q \not \in {\mathcal{C}}$}}
\end{array} \right.  \label{L1def} \\
\lefteqn{ L_2(q,b,c) \;=\; \frac{i \qslsh}{32 \pi^3} } \nonumber \\
&&\times \left\{ \begin{array}{cl}
\displaystyle \sqrt{\Delta} \:\frac{a+b-c}{2a^2}\:
\Theta(\sqrt{b} - \sqrt{a} - \sqrt{c}) -
\frac{(b-c)^2-2ac}{2a^2}\: \Theta(b-c)
& {\mbox{if $q \in {\mathcal{C}}^\lor$}} \\[.8em]
\displaystyle \sqrt{\Delta} \:\frac{a+b-c}{2a^2}\:
\Theta(\sqrt{c} - \sqrt{a} - \sqrt{b}) +
\frac{(b-c)^2-2ac}{2a^2}\: \Theta(c-b)
& {\mbox{if $q \in {\mathcal{C}}^\land$}} \\[.8em]
\displaystyle \sqrt{\Delta} \:\frac{a+b-c}{4a^2}
- \frac{(b-c)^2-2ac}{4a^2}\:\epsilon(b-c)
& {\mbox{if $q \not \in {\mathcal{C}}$}}\:, \label{L2def}
\end{array} \right.
\end{eqnarray}
where again~$a=q^2$.
\end{Lemma}
{\Proof} Using that differentiation in position space corresponds to
multiplication in momentum space and vice versa, we obtain from~(\ref{r1})
that
\begin{eqnarray*}
\Pdd_\xi f^\varepsilon(k) &=& i k\slsh\: f(k^2)\: e^{\varepsilon k^0}
\;=\; \frac{i}{2}\: (\Pdd_k h(k^2))\: e^{\varepsilon k^0} \\
&=& \frac{i}{2}\: \Pdd_k \left(h(k^2)\: e^{\varepsilon k^0} \right) \:-\:
\frac{i \varepsilon}{2}\: \gamma^0\: h(k^2)\: e^{\varepsilon k^0} 
\;=\; \frac{1}{2} \left(-\xi\slsh - i \varepsilon \gamma^0 \right)
h^\varepsilon(k) \:,
\end{eqnarray*}
where the negative, regularized distribution $h^\varepsilon$ is defined by
\[ h^\varepsilon(k) \;=\; h(k^2)\: e^{\varepsilon k^0} \qquad
{\mbox{and}} \qquad h(k^2) \;=\; \int_0^{k^2} f(e)\: de\:. \]
Now we can compute the product $\Pdd f \cdot \overline{g}$ with
the help of Lemma~\ref{lemma33},
\begin{eqnarray*}
\Pdd f \cdot \overline{g} &=& \lim_{\varepsilon \searrow 0}
\frac{1}{2} \left(-\xi\slsh - i \varepsilon \gamma^0 \right) h^\varepsilon
\: \overline{g^\varepsilon} \\
&=& {\mbox{(l.c.)}} + \int_0^\infty dc\: h(c) \int_0^\infty db\: g(b)\:
\frac{-\xi\slsh}{2}\: K_1(q,b,c)\:.
\end{eqnarray*}
Applying the transformations $\xi\slsh = -i \Pdd_q$ and
\[ \int_0^\infty dc\:h(c)\: K_1(q,b,c) \;=\;
\int_0^\infty dc \int_0^c de\: f(e)\: K(q,b,c)
\;=\; \int_0^\infty de\: f(e) \int_e^\infty dc\: K_1(q,b,c)\:, \]
we obtain the representation~(\ref{s1s}) with
\[ L_1(q,b,e) \;=\; \int_e^\infty \frac{i}{2}\: \Pdd_q K_1(q,b,c)\:dc\:. \]

In order to compute $L_1$ more explicitly, we substitute for $K_1$ the
expression~(\ref{K1def}) and carry out the $c$-integration.
The following indefinite integrals are easily verified by differentiation,
\begin{eqnarray*}
\int \frac{d}{da} \frac{\sqrt{\Delta}}{a}\: dc &=& \sqrt{\Delta}\;
\frac{b-c-a}{2a^2} \\
\int \frac{d}{da} \frac{|b-c|}{a}\: dc &=& |b-c|\:
\frac{b-c}{2a^2} \:.
\end{eqnarray*}
The asymptotic expansion for large $c$
\begin{eqnarray*}
\sqrt{\Delta} &=& (c-b) \:-\: a\: \frac{c+b}{c-b} \:+\: {\cal{O}}(c^{-2}) \\
\sqrt{\Delta}\: (b-c-a) &=& -(c-b)^2 + 2 ab + a^2 + {\cal{O}}(c^{-1})
\end{eqnarray*}
shows that the difference of the two indefinite integrals has a finite
limit as $c \to \infty$,
\[ \lim_{c \to \infty} \left(
\sqrt{\Delta}\:\frac{b-c-a}{2a^2} - |b-c|\: \frac{b-c}{2a^2} \right)
\;=\; \frac{b}{a} + \frac{1}{2} \:. \]
Taking into account the boundary conditions as determined by the
Heaviside functions in~(\ref{K1def}) gives for $L_1$ the expression
\[ \frac{i \qslsh}{32\pi^3}\left\{ \begin{array}{cl}
\displaystyle \sqrt{\Delta} \:\frac{a-b+c}{2a^2}\:
\Theta(\sqrt{b} - \sqrt{a} - \sqrt{c}) +
\frac{(b-c)^2}{2a^2}\: \Theta(b-c)
& {\mbox{if $q \in {\mathcal{C}}^\lor$}} \\[.8em]
\displaystyle \sqrt{\Delta} \:\frac{a-b+c}{2a^2}\:
\Theta(\sqrt{c} - \sqrt{a} - \sqrt{b}) -
\frac{(b-c)^2}{2a^2}\: \Theta(c-b) + \frac{b}{a} + \frac{1}{2}
& {\mbox{if $q \in {\mathcal{C}}^\land$}} \\[.8em]
\displaystyle \sqrt{\Delta} \:\frac{a-b+c}{4a^2}\: +
\frac{|b-c|\: (b-c)}{4a^2} + \frac{b}{2a} + \frac{1}{4}
& {\mbox{if $q \not \in {\mathcal{C}}$}}\:.
\end{array} \right. \]
We finally subtract the distribution
\[ b\:\Theta(b-c)\: \frac{\mbox{PP}}{a}\:+\: \frac{1}{4}\: \Theta(a)\: \epsilon(-q^0)
+ \frac{1}{4} \:, \]
which can be written in the form~(\ref{harmonic}) and is
therefore localized on the light cone. This gives~(\ref{L1def}).

The formula for $L_2$, (\ref{L2def}), is obtained from~(\ref{L1def}) by
double conjugation, $f \cdot \Pdd \overline{g} =
\Pdd \overline{g} \cdot \overline{\overline{f}}$. This gives the identity
$L_2(q,b,c) = L_1(-q,c,b)$.
\QED
It is instructive to verify the Leibniz rule
\[ \Pdd (f \cdot \overline{g}) - (\Pdd f \cdot \overline{g}) - (f \cdot \Pdd \overline{g})
\;=\; 0 \]
by substituting the results of Lemma~\ref{lemma33} and Lemma~\ref{lemma34}.
This gives
\[ \left( \Pdd (f \cdot \overline{g}) - (\Pdd f \cdot \overline{g})
- (f \cdot \Pdd \overline{g}) \right)(q) \;=\; {\mbox{\rm{(l.c.)}}} +
\int_0^\infty dc\: f(c) \int_0^\infty db\: g(b) \: M(q,b,c) \]
with
\begin{eqnarray*}
M(q,b,c) &=& i \qslsh\: K_1(q,b,c) - L_1(q,b,c) - L_2(q,b,c) \\
&=& \frac{i \qslsh}{32 \pi^3}
\left\{ \begin{array}{cl}\displaystyle 
- \frac{|b-c|}{a}\: \Theta(b-c) + \frac{b-c}{2a}\: \Theta(b-c) &
{\mbox{if $q \in {\mathcal{C}}^\lor$}} \\[.8em] \displaystyle 
- \frac{|b-c|}{a}\: \Theta(c-b) - \frac{b-c}{a}\: \Theta(c-b) &
{\mbox{if $q \in {\mathcal{C}}^\land$}} \\[.8em] \displaystyle
-\frac{|b-c|}{2a} + \frac{b-c}{2a}\: \epsilon(b-c) &
{\mbox{if $q \not \in {\mathcal{C}}$}} \end{array} \right. \\
&=& 0\:.
\end{eqnarray*}

\subsection{Convolutions Involving Dirac Seas} \label{sechM}
We now apply the previous results on convolution integrals to composite expressions
in the fermionic projector. We begin with the convolution integral~(\ref{cFdef}).
\begin{Thm} \label{thm57}
For any~$k \in {\mathcal{C}}$ with~$k^2>0$,
\beq \label{convrep}
\hF(k) \;=\;
\frac{1}{64\, \pi^3}\: \frac{k\slsh}{k^4}\, \epsilon(k^0)
\sum_{\alpha, \beta=1}^g \rho_\alpha \,\rho_\beta\:
J(k^2, m_\alpha, m_\beta)\:,
\eeq
where
\beq \label{Jdef}
J(a,x,y) \;=\; -\sqrt{\Delta(a,x^2,y^2)}\:(x-y)\, \epsilon(x^2-y^2)  \left[(x+y)^2 -a\right]
\Theta \!\left((|x|-|y|)^2 - a \right)
\eeq
with~$\Delta$ according to~(\ref{defDelta}).
\end{Thm}
{\Proof} Introducing the distributions
\beq \label{fgdef}
f(a) \;=\; \sum_{\beta=1}^g \rho_\beta\: \delta(a-m_\beta^2) \qquad {\mbox{and}} \qquad
g(a) \;=\; \sum_{\beta=1}^g \rho_\beta\:m_\beta\: \delta(a-m_\beta^2)\:,
\eeq
we can write the fermionic projector~(\ref{hPdef}) with the notation~(\ref{rules})
as~$P = -i \Pdd f + g$. Then
\beq \label{PbP}
(P \cdot \overline{P})_0 \;=\; -i \left( (\Pdd f) \cdot \overline{g}
- g \cdot (\Pdd \overline{f}) \right) .
\eeq
For~$k \in C^\wedge$, the appearing convolution integrals were computed in
Lemma~\ref{lemmamix}. We thus obtain
\begin{eqnarray*}
(P \cdot \overline{P})_0 &=& -i \Pdd \alpha(a) \, \\
\alpha(a) &=& \frac{1}{32 \pi^3} \int_0^\infty dc\: f(c) \int_0^\infty
db \:g(b)\:\sqrt{\Delta}\; \frac{a-b+c}{2 a^2} \\
&& \spc \times \left( \Theta(\sqrt{c}-\sqrt{a}-\sqrt{b}) - \Theta(\sqrt{b}-\sqrt{a}-\sqrt{c}) \right) ,
\end{eqnarray*}
where we set~$a=k^2$. Using~(\ref{fgdef}), we can carry out the integrals to
obtain~(\ref{convrep}) with
\begin{eqnarray*}
J(a,x,y) &=& -2 x\, \sqrt{\Delta(a,x^2,y^2)}\; (a-x^2+y^2)
\left( \Theta(|y|-\sqrt{a}-|x|) - \Theta(|x|-\sqrt{a}-|y|) \right) \\
&=& -2 x\, \sqrt{\Delta(a,x^2,y^2)}\; (a-x^2+y^2) \: \epsilon(|y|-|x|)\:
\Theta\!\left( (|y|-|x|)^2-a \right) .
\end{eqnarray*}
As~(\ref{convrep}) is symmetric in~$\alpha$ and~$\beta$, we may symmetrize~$J$
in~$x$ and~$y$, giving~(\ref{Jdef}). For~$q \in {\mathcal{C}}^\vee$, the result
follows similarly using the remark after the proof of Lemma~\ref{lemmamix}.
\QED

Next we can introduce the distribution~$\hM$, which already appeared in our
sketch as the Fourier transform of the distribution~$\tM$ defined by (\ref{Mtdef}).
Here we proceed in the opposite way: we first define~$\tM$ and then show that
its Fourier transform really satisfies~(\ref{Mtdef}). We denote this Fourier
transform without a tilde by~$\M$. This clarifies that~$\M$ is a specific distribution
satisfying~(\ref{Mtdef}), whereas~$\tM$ is not unique but can be changed
freely according to~(\ref{tMmod}).
\begin{Thm}  \label{thm58}
The function
\beq
\hM(k) \;=\; \frac{1}{64 \pi^3} \, \frac{k\slsh}{k^4}\: \Theta(k^2)\, \epsilon(k^0)
\sum_{\alpha, \beta=1}^g \rho_\alpha \,\rho_\beta \left(
J(k^2, m_\alpha, m_\beta) \;+\; K(k^2, m_\alpha, m_\beta) \right) \qquad \label{e1}
\eeq
with~$J$ according to~(\ref{Jdef}) and
\beq
K(a, x, y) \;=\; (x-y)^2 (x+y)^3 \:-\: 2 a\, (x^3+y^3) \label{Kdef}
\eeq
defines a tempered distribution, if the pole on the cone~$\{k \:|\: k^2=0\}$ is
understood as the distributional derivative of a logarithm, i.e.
\beq \label{interpret}
\frac{k\slsh}{k^2}\,\Theta(k^2)\, \epsilon(k^0) \qquad {\mbox{stands for}} \qquad
\frac{1}{2}\: \Pdd_k \!\left(\log(k^2) \, \Theta(k^2)\,\epsilon(k^0)  \right) .
\eeq
Its Fourier transform~$\M(\xi)$ satisfies away from the light cone the relation
\beq \label{tMc}
\tM(\xi) \;=\; 2 A_0(\xi) \qquad \forall \xi \not \in L
\eeq
with~$A_0$ as given by~(\ref{Adef}).
\end{Thm}
{\Proof} It suffices to consider any pair~$\alpha, \beta \in \{1,\ldots, g\}$.
In the case~$m_\alpha \neq m_\beta$, we write out~(\ref{tMc})
in momentum space to obtain similar to~(\ref{PbP}) the condition
\[ \hM(k) \;=\; -2i \left( (\Pdd f) \cdot \overline{g}
- g \cdot (\Pdd \overline{f}) \right) \:. \]
Computing the convolution integrals with the help of Lemma~\ref{lemma34} modulo singular
contributions on the light cone, a straightforward calculation similar to that
in the proof of Theorem~\ref{thm1} gives~(\ref{e1}) with~$K$ according to~(\ref{Kdef}).
Expanding the factor~$\sqrt{\Delta}$ in~(\ref{Jdef}) for small~$a$,
\beq \label{Delex0}
\sqrt{\Delta(a,b,c)} \;=\; |b-c| - a \: \frac{b+c}{|b-c|} + {\cal{O}}(a^2)\:,
\eeq
one sees that~$\hM(k)$ has no poles on the cone~$\{ k \:|\: k^2=0\}$.
Thus~$\hM(k)$ is a $L^1_{\mbox{\scriptsize{loc}}}$-function, which clearly
defines a tempered distribution.

In the case~$m_\alpha=m_\beta$, the function~$J$ vanishes identically, and thus
\[ \hM(k) \;=\; -\frac{1}{16 \pi^3} \, \frac{k\slsh}{k^2}\, \epsilon(k^0)\: \Theta(k^2)\;
\rho_\alpha \,\rho_\beta \: m_\alpha^3 . \]
Interpreting the pole according to~(\ref{interpret}), this is again a tempered distribution.
It has such a simple form that~(\ref{tMc}) can be verified by explicit
calculation. Namely, the Fourier transform of~$\hM$ can be given in closed form
(see Corollary~\ref{lemmaA2} below). Comparing this formula with an
explicit computation of~$A_0$ in position shows that~(\ref{tMc}) is indeed satisfied.
\hspace*{1cm} \QED
Let us briefly discuss the result of Theorem~\ref{thm57} and Theorem~\ref{thm58}.
We first point out that, due to the Heaviside function in~(\ref{Jdef}), the
function~$\hF(k)$ vanishes identically for large~$k^2$,
\beq \label{Fvanish}
\hF(k) \;=\; 0 \qquad {\mbox{if~$k^2 > \max_{\beta \in \{1,\ldots, g\}} m_\beta^2$}} \:.
\eeq
However, it has poles as~$k^2 \searrow 0$, as one sees from the expansion~(\ref{Delex0}),
\begin{eqnarray}
\hF(k)
&=&  -\:\frac{ k\slsh\, \epsilon(k^0)}{64\, \pi^3} \!\!\!
\sum_{\alpha, \beta {\mbox{\scriptsize{ with }}} m_\alpha \neq m_\beta} \!\!\!
\rho_\alpha \,\rho_\beta
\left(\frac{(m_\alpha-m_\beta)^2 (m_\alpha+m_\beta)^3}{k^4} -
2\: \frac{m_\alpha^3+m_\beta^3}{k^2} \right) \nonumber \\
&&+ k\slsh \:{\mathcal{O}}(k^0) \:. \label{convex}
\end{eqnarray}
The distribution~$\hM$ differs from~$\hF$
by the function~$K$ in~(\ref{e1}) and by
the fact that the poles of~$\hM$
have a meaning in the distributional sense~(\ref{interpret}).
The order of the pole is smaller than that in~(\ref{convex}),
\beq \label{hMex}
\hM(k) \;=\; -\frac{1}{32 \pi^3} \, k\slsh\: \Theta(k^2)\, \epsilon(k^0)\!\!\!
\sum_{\alpha, \beta {\mbox{\scriptsize{ with }}} m_\alpha = m_\beta} \!\!\!
\rho_\alpha \,\rho_\beta \:\frac{m_\alpha^3+m_\beta^3}{k^2} \:+\: k\slsh \:{\mathcal{O}}(k^0)\:.
\eeq
Since~(\ref{convrep}) vanishes identically for large~$k^2$, 
using~(\ref{m3def}, \ref{m5def}) we find that~$\hM$ has the simple asymptotics
\beq \label{hMexinf}
\hM(k) \;=\; 2\pi^2 \:k\slsh\:\Theta(k^2)\, \epsilon(k^0)
\left(\frac{\m_3}{k^2} + \frac{4 \m_5}{k^4} \right) \qquad
 {\mbox{if~$k^2 > \max_{\beta \in \{1,\ldots, g\}} m_\beta^2$}}\:.
\eeq

We end this section by specifying the effect of the parameters~$c_0$ and~$c_1$
in~(\ref{tMmod}) on the integral in~(\ref{dSrel1}).
\begin{Lemma} \label{lemmac0c1comp}
The Fourier transforms of the distributions
\[ A(\xi) \;:=\; \xi\slsh\, \delta(\xi^2)\: \epsilon(\xi^0) \:,\qquad
B(\xi) \;:=\; \xi\slsh\, \delta'(\xi^2)\: \epsilon(\xi^0) \]
satisfy the relations
\begin{eqnarray}
\int_0^\infty \Tr \left( (\hat{A}*\hat{P})(q)  \: \delta \hat{P}(q) \right)  a\: da &=&
32 \pi^4\, \delta \m_3 \label{m3var} \\
\int_0^\infty \Tr \left( (\hat{B}*\hat{P})(q)  \: \delta \hat{P}(q) \right)  a\: da &=&
-32 \pi^4\, \delta \m_5 \:, \label{m5var}
\end{eqnarray}
where~$\m_3$ and~$\m_5$ are defined by~(\ref{m3def}, \ref{m5def}), and where again~$a=q^2$.
\end{Lemma}
{\Proof} Again using the notation~(\ref{rules}), we write the fermionic projector
as~$\hat{P} = -i \Pdd f + g$ with~$f$ and~$g$ according to~(\ref{fgdef}).
The Fourier transforms of~$A$ and~$B$ can be computed explicitly (see Corollary~\ref{lemmaA1}
below). Decomposing them into a sum of a positive and a negative distribution, we write~$\hat{A}$
and~$\hat{B}$ as
\[ i \Pdd (h(a) - \overline{h}(a)) \]
with
\[ h(a) \;=\; 8 \pi^2\, \delta'(a) \;\; {\mbox{for~$\hat{A}$}} \qquad {\mbox{and}} \qquad
h(a) \;=\; 2 \pi^2\, \delta(a) \;\; {\mbox{for~$\hat{B}$}} \:. \]
The convolutions~$\hat{P} * \hat{A}$ and~$\hat{P} * \hat{B}$ are then expressed by
\[ (\Pdd f + i g) \cdot (\Pdd h - \Pdd \overline{h}) . \]
For any~$q \in {\mathcal{C}}^\wedge$, we can write these convolutions with the help of
Lemma~\ref{lemmapos} and Lemma~\ref{lemmamix},
\begin{eqnarray*}
(\Pdd f \cdot (\Pdd h - \Pdd \overline{h}))(a) &=& \alpha(a)\:, \\
\alpha(a) &=& \frac{1}{32 \pi^3} \int_0^\infty dc \:f(c)\: \epsilon(a-c) \int_0^{(\sqrt{a}
-\sqrt{c})^2} db \:h(b)\:\sqrt{\Delta}\; \frac{c+b-a}{2 a} \\
(i g \cdot (\Pdd h - \Pdd \overline{h}))(a) &=& i \Pdd \beta(a)\:, \\
\beta(a) &=& \frac{1}{32 \pi^3} \int_0^\infty dc \:g(c)\, \epsilon(a-c) \int_0^{(\sqrt{a}
-\sqrt{c})^2} db \:h(b)\:\sqrt{\Delta}\; \frac{a+b-c}{2 a^2}\:.
\end{eqnarray*}

We now compute the remaining integrals using the special form of the functions~$f$,
$g$ and~$h$. We begin with the calculation for~$\hat{A}$. Using similar
to~(\ref{Delex0}) the expansion
\beq \label{Delex}
\sqrt{\Delta} \;=\; |a-c| - b \: \frac{a+c}{|a-c|} + {\cal{O}}(b^2)\:,
\eeq
we obtain
\begin{eqnarray*}
\alpha(a) &=&  -\frac{1}{4 \pi} \int_0^\infty dc \:f(c)\: \epsilon(a-c) 
\frac{d}{db} \!\left( \sqrt{\Delta}\; \frac{c+b-a}{2 a} \right) \Big|_{b=0} \\
&=& -\frac{1}{4 \pi} \int_0^\infty dc \:f(c)
\;=\; -\frac{1}{4 \pi} \sum_{\alpha=1}^g \rho_\alpha \\
\beta(a) &=&  -\frac{1}{4 \pi} \int_0^\infty dc \:g(c)\: \epsilon(a-c) 
\frac{d}{db} \!\left( \sqrt{\Delta}\; \frac{a+b-c}{2 a^2} \right) \Big|_{b=0} \\
&=& \frac{1}{4 \pi} \int_0^\infty dc \:g(c)\; \frac{c}{a^2}
\;=\; \frac{1}{4 \pi} \sum_{\alpha=1}^g \rho_\alpha\: \frac{m_\alpha^3}{a^2}
\end{eqnarray*}
and thus
\[ (\hat{P}*\hat{A})(q) \;=\;
-\frac{1}{4 \pi} \sum_{\alpha=1}^g \: \frac{\rho_\alpha}{a^2} 
\left(m_\alpha q\slsh + a \right) \:. \]
Now we take the trace with a ``test Dirac sea''
\[ \hat{R}(q) \;:=\; \rho \,(q\slsh+m)\, \delta(q^2-m^2) \:. \]
This gives
\beq \label{m3s}
\int_0^\infty \Tr \left( (\hat{P}*\hat{A})(q)  \:\hat{R}(q) \right)  a\: da
\;=\; -\frac{1}{\pi} \sum_{\alpha=1}^g \rho\, \rho_\alpha\: 
\left(m_\alpha^3 + m^3 \right) .
\eeq
After varying~$\rho$ and~$m$, we can build up~$\delta \hat{P}$ by taking the sum of~$g$
test Dirac seas. Using the symmetry of~(\ref{m3s}) in~$\rho, \rho_\alpha$ and~$m, m_\alpha$,
we thus obtain
\[ \int_0^\infty \Tr \left( (\hat{P}*\hat{A})(q)  \:\delta \hat{P}(q) \right)  a\: da
\;=\; -\frac{1}{2 \pi} \;\delta \!\left(\sum_{\alpha=1}^g \rho_\alpha\, \rho_\beta\: 
\left(m_\alpha^3 + m_\beta^3 \right) \right) , \]
and comparing with~(\ref{m3def}) gives~(\ref{m3var}).

The calculation for~$\hat{B}$ is similar. Using~(\ref{Delex}), we obtain
\begin{eqnarray*}
\alpha(a) &=&  \frac{1}{16 \pi} \int_0^\infty dc \:f(c)\: \epsilon(a-c) 
\sqrt{\Delta}\; \frac{c+b-a}{2 a} \Big|_{b=0} \\
&=& -\frac{1}{16 \pi} \int_0^\infty dc \:f(c)\; \frac{(c-a)^2}{2 a}
\;=\; -\frac{1}{32 \pi} \sum_{\alpha=1}^g \rho_\alpha\: \frac{(m_\alpha^2-a)^2}{a} \\
\beta(a) &=&  \frac{1}{16 \pi} \int_0^\infty dc \:g(c)\; \frac{(c-a)^2}{2 a^2}
\;=\;\frac{1}{32 \pi} \sum_{\alpha=1}^g \rho_\alpha\: \frac{m_\alpha \,(m_\alpha^2-a)^2}{a^2}
\end{eqnarray*}
and thus
\begin{eqnarray*}
(\hat{P}*\hat{B})(q) &=& -\frac{1}{32 \pi} \sum_{\alpha=1}^g \rho_\alpha\:
\frac{(m_\alpha^2-a)^2}{a^2}  \left(m_\alpha q\slsh + a \right) \\
\int_0^\infty \Tr \left( (\hat{P}*\hat{B})(q)  \:\hat{R}(q) \right)  a\: da
&=& -\frac{1}{8 \pi} \sum_{\alpha=1}^g \rho\, \rho_\alpha\: (m_\alpha-m)^2
\left(m_\alpha + m \right)^3.
\end{eqnarray*}
Comparing with~(\ref{m5def}) gives~(\ref{m5var}).
\QED

\section{A Lorentz Invariant Regularization and its Fourier Transform}
\setcounter{equation}{0}  \label{sec6}
According to~(\ref{Mtdef}) and~(\ref{tA}, \ref{fexp}), on the light cone $\tM$ has
the leading pole
\beq \label{leading}
\tM(\xi) \;=\; \m_3\: \frac{\xi\slsh}{\xi^4}\, \Theta(\xi^2)\,\epsilon(\xi^0)
\:+\: {\mathcal{O}}(\xi^{-2}) \:,
\eeq
and this pole needs to be regularized in order
to give the action a mathematical meaning. In this section we shall construct a
Lorentz invariant regularization of the pole in~(\ref{leading}), which is explicit
both in position and in momentum space.
Our analysis is based on the following Fourier integral.
\begin{Lemma} \label{lemma61}
The following identity holds in the sense of distributions:
\beq \label{eq61}
\int \frac{d^4k}{(2 \pi)^4}\: e^{-\frac{\varepsilon k^2}{2}}\: \Theta(k^2)\, \epsilon(k^0)\:
e^{i k \xi} \;=\; -\frac{i}{4 \pi^2 \varepsilon^2} \:e^{-\frac{\xi^2}{2\varepsilon}}\:
\Theta(\xi^2) \,\epsilon(\xi^0) \:+\: \frac{i}{2 \pi^2 \varepsilon} \:
\delta(\xi^2) \,\epsilon(\xi^0)
\eeq
\end{Lemma}
{\Proof} Due to spherical symmetry, we can assume that~$\xi=(t,r,0,0)$ with~$r \geq 0$.
Choosing polar coordinates~$k=(\omega, p \cos \vartheta, p \sin \vartheta \cos \varphi,
p \sin \vartheta \sin \varphi)$, the Fourier integral becomes
\begin{eqnarray*}
A &:=& \int \frac{d^4k}{(2 \pi)^4}\: e^{-\frac{\varepsilon k^2}{2}}\: \Theta(k^2)\, \epsilon(k^0)\:
e^{i k \xi} \\
&=& \frac{1}{8 \pi^3} \int_{-\infty}^\infty d\omega\, \epsilon(\omega)\, e^{i \omega t}
\int_0^{|\omega|} p^2\, dp\: e^{-\frac{\varepsilon}{2}\left( \omega^2-p^2 \right)}
\int_{-1}^1 d\cos \vartheta\; e^{-i p r\, \cos \vartheta} \\
&=& \frac{i}{8 \pi^3 r} \int_{-\infty}^\infty d\omega\, \epsilon(\omega)\, e^{i \omega t}
\int_0^{|\omega|} p\, dp\: e^{-\frac{\varepsilon}{2}\left( \omega^2-p^2 \right)}
\left( e^{-i p r} - e^{ipr} \right) \\
&=& \frac{i}{8 \pi^3 r} \int_{-\infty}^\infty d\omega\, \epsilon(\omega)\,
\int_{-|\omega|}^{|\omega|} p\, dp\: e^{-\frac{\varepsilon}{2}\left( \omega^2-p^2 \right)
+i (\omega t-p r)}\:.
\end{eqnarray*}
We now introduce the ``mass cone coordinates''
\[ u \;=\; \frac{1}{2}\: (\omega+p) \:,\qquad v \;=\; \frac{1}{2}\: (\omega-p)\:, \]
and light cone coordinates
\[ s \;=\; \frac{1}{2}\: (t-r) \:,\qquad\; l \;=\; \frac{1}{2}\: (t+r)\:. \]
Then our integrals transform to
\[ A \;=\; \frac{i}{4 \pi^3 r} \left( \int_0^\infty \int_0^\infty - 
\int_{-\infty}^0 \int_{-\infty}^0 \right) du\, dv\; (u-v)\: e^{-2 \varepsilon u v + 2 i u s + 2 i v l}\:. \]
Carrying out the $v$-integral in the case~$u>0$ gives
\[ \int_0^\infty (u-v) \:e^{-2 \varepsilon u v + 2 i v l} \: dv \;=\;
\left(u+ \frac{i}{2}\, \frac{\partial}{\partial l} \right)
\int_0^\infty e^{-2 \varepsilon u v + 2 i v l} \: dv
\;=\; \left(u+ \frac{i}{2}\, \frac{\partial}{\partial l} \right) \!\frac{1}{2 \varepsilon u - 2 i l} \:, \]
whereas in the case~$u<0$ we get the same expression with the opposite sign.
Hence
\begin{eqnarray*}
A &=& \frac{i}{4 \pi^3 r} \int_{-\infty}^\infty \left[
\left(u+ \frac{i}{2}\, \frac{\partial}{\partial l} \right) \!\frac{1}{2 \varepsilon u - 2 i l} \right]
e^{2 i u s} \, du \\
&=& \frac{1}{8 \pi^3 r} \left(\frac{\partial}{\partial s} - \frac{\partial}{\partial l} \right)
\int_{-\infty}^\infty \frac{1}{2 \varepsilon u - 2 i l} \:e^{2 i u s} \, du \\
&=& \frac{i}{8 \pi^2 r \varepsilon} \left(\frac{\partial}{\partial s} - \frac{\partial}{\partial l} \right)
\left[ \epsilon(s)\: \Theta(s l)\, e^{-\frac{2 s l}{\varepsilon}} \right] ,
\end{eqnarray*}
where in the last step we computed the integral with residues. We finally transform
back to polar coordinates,
\begin{eqnarray*}
A &=& -\frac{i}{4 \pi^2 \varepsilon} \;\frac{1}{r} \frac{\partial}{\partial r}
\left[ \epsilon(t)\: \Theta(t^2-r^2)\, e^{-\frac{t^2-r^2}{2 \varepsilon}} \right] ,
\end{eqnarray*}
carry out the distributional derivatives and write the result in Lorentz invariant form.
\QED
Before going on, we collect of a few simple Fourier integrals which can
easily be obtained from~(\ref{eq61}).
\begin{Corollary} \label{lemmaA1}
The following equations hold in the sense of distributions:
\begin{eqnarray}
\int \delta'(\xi^2)\, \epsilon(\xi^0)\; e^{-i k \xi}\, d^4 \xi &=& -i \pi^2\:
\Theta(k^2)\, \epsilon(k^0) \label{A14} \\
\int \delta(\xi^2)\, \epsilon(\xi^0)\; e^{-i k \xi}\, d^4 \xi &=& -4i \pi^2\:
\delta(k^2)\, \epsilon(k^0) \label{A11} \\
\int \xi\slsh\, \delta'(\xi^2)\, \epsilon(\xi^0)\; e^{-i k \xi}\, d^4 \xi &=& 2 \pi^2\,
k\slsh\, \delta(k^2)\, \epsilon(k^0) \label{A13} \\
\int \xi\slsh \,\delta(\xi^2)\, \epsilon(\xi^0)\; e^{-i k \xi}\, d^4 \xi &=& 8 \pi^2\,
k\slsh\, \delta'(k^2)\, \epsilon(k^0) \label{A12}
\end{eqnarray}
\end{Corollary}
{\Proof} The identity~(\ref{A14}) could be obtained by a direct computation
of the Fourier integrals. We here use another method where we recover~(\ref{A14})
as a limiting case of Lemma~\ref{lemma61}. Using that
\[ \frac{1}{\varepsilon^2} \int_0^\infty e^{-\frac{z}{2 \varepsilon}}\, dz \:-\:
\frac{2}{\varepsilon} \;=\; 0 \:,\qquad
 \frac{1}{\varepsilon^2} \int_0^\infty z\, e^{-\frac{z}{2 \varepsilon}}\, dz 
 \;=\; 4 \:, \]
one sees that
\beq \label{vorpla}
\frac{1}{\varepsilon^2} \:e^{-\frac{z}{2\varepsilon}}\, \Theta(z)
\:-\: \frac{1}{2 \varepsilon} \: \delta(z) 
\;\stackrel{\varepsilon \rightarrow 0}{\longrightarrow}\; -4\, \delta'(z)
\eeq
with convergence in~${\mathcal{S}}'(\R)$. This also shows
that~(\ref{eq61}) converges in~${\mathcal{S}}'(\hat{M})$ to
\[ \int \frac{d^4k}{(2 \pi)^4}\: \Theta(k^2)\, \epsilon(k^0)\:
e^{i k \xi} \;=\; \frac{i}{\pi^2} \:\delta'(\xi^2) \,\epsilon(\xi^0)\:, \]
except maybe at the origin~$\xi=0$, where the light cone has its cusp.
To prove convergence at the origin, one verifies easily by a
scaling argument that for any test function~$\eta \in C^\infty_0(M)$,
\beq \label{origin}
\lim_{\delta \searrow 0} \:\lim_{\varepsilon \searrow 0}
\int \eta\!\left( \frac{\xi}{\delta} \right) \left[
\frac{1}{\varepsilon^2} \:e^{-\frac{\xi^2}{2\varepsilon}}\:
\Theta(\xi^2) \,\epsilon(\xi^0) \:-\: \frac{1}{2 \varepsilon} \:
\delta(\xi^2) \,\epsilon(\xi^0) \right] \;=\; 0\:.
\eeq
We finally apply Plancherel's theorem to~(\ref{vorpla}) to obtain~(\ref{A14}).

Applying the operator~$\Box_k$ to~(\ref{A14}) gives~(\ref{A11}).
To derive~(\ref{A13}), we apply the operator~$i \Pdd_k$ to~(\ref{A14}).
Finally, the identity~(\ref{A12}) follows from~(\ref{A11}) by applying the
differential operator~$i \Pdd_k$. 
\QED

The main importance of Lemma~\ref{lemma61} lies in the fact that integrating over~$\varepsilon$
gives the Fourier transformations of regularized poles.
\begin{Prp} The following equations hold in the sense of distributions:
\begin{eqnarray}
\lefteqn{ \int \frac{d^4k}{(2 \pi)^4}\: e^{-\frac{\varepsilon k^2}{2}}\:\Box_k \!\left( \log(k^2)\, \Theta(k^2)\, \epsilon(k^0) \right)
e^{i k \xi} } \nonumber \\
&=&-\frac{i}{\pi^2} \left[ \frac{1}{\xi^2} \left(1 -  e^{-\frac{\xi^2}{2\varepsilon}} \right)
\Theta(\xi^2) \,\epsilon(\xi^0)
\:+\: \delta(\xi^2) \,\epsilon(\xi^0)\, (c-1+\log \varepsilon) \right]
\label{611} \\
\lefteqn{ \int \frac{d^4k}{(2 \pi)^4} \:e^{-\frac{\varepsilon k^2}{2}}\: \Pdd_k \!\left( \log(k^2)\, \Theta(k^2)\, \epsilon(k^0) \right)
e^{i k \xi} } \nonumber \\
&=& -\frac{1}{2 \pi^2}\: \Pdd_\xi\!
\left[ \frac{1}{\xi^2} \left(1 -  e^{-\frac{\xi^2}{2\varepsilon}} \right)
\Theta(\xi^2) \,\epsilon(\xi^0)
\:+\: \delta(\xi^2) \,\epsilon(\xi^0)\, (c+\log \varepsilon) \right] \label{612}
\end{eqnarray}
Here the constant~$c$ is given by
\beq \label{cdef}
c \;=\; \gamma - \log2\:,
\eeq
and~$\gamma$ is Euler's constant.
\end{Prp}
{\Proof}
Integrating the variable~$\varepsilon$ in~(\ref{eq61}) over the compact interval~$[\varepsilon, L]$
with~$L > \varepsilon$ gives the identity
\begin{eqnarray*}
\lefteqn{ \int \frac{d^4k}{(2 \pi)^4}\: \frac{2}{k^2} \left(e^{-\frac{\varepsilon k^2}{2}}
- e^{-\frac{L k^2}{2}} \right)\: \Theta(k^2)\, \epsilon(k^0)\: e^{i k \xi}} \\
&=& -\frac{i}{4 \pi^2} \:\frac{2}{\xi^2} \left(e^{-\frac{k^2}{2L}}
-  e^{-\frac{\xi^2}{2\varepsilon}} \right)
\Theta(\xi^2) \,\epsilon(\xi^0)
\:+\: \frac{i}{2 \pi^2} \:\delta(\xi^2) \,\epsilon(\xi^0) \left(\log L - \log \varepsilon \right) .
\end{eqnarray*}
Bringing the term involving~$\log L$ to the left,
we can use~(\ref{A11}) to combine it with the integrand,
\begin{eqnarray}
\lefteqn{ \int \frac{d^4k}{(2 \pi)^4} \left\{ \frac{1}{k^2} \left(e^{-\frac{\varepsilon k^2}{2}}
- e^{-\frac{L k^2}{2}} \right)\: \Theta(k^2)\, \epsilon(k^0)
\:-\: \delta(k^2)\, \epsilon(k^0) \:(c+\log L) \right\}e^{i k \xi}} \nonumber \\
&=& -\frac{i}{4 \pi^2} \left[ \frac{1}{\xi^2} \left(e^{-\frac{k^2}{2L}}
-  e^{-\frac{\xi^2}{2\varepsilon}} \right)
\Theta(\xi^2) \,\epsilon(\xi^0)
\:+\: \delta(\xi^2) \,\epsilon(\xi^0)\, (c+\log \varepsilon) \right] . \label{inid}
\end{eqnarray}
A direct calculation yields
for any~$\eta \in C^\infty_0(\R)$ and~$x>0$ the relations
\begin{eqnarray}
\lefteqn{ \lim_{L \rightarrow \infty}
\left( \int_{-\infty}^x \eta(a)\: \frac{1}{a} \left((1 - e^{-\frac{La}{2}})\, \Theta(a) \right) da\:-\: \log L 
\right) } \nonumber \\
&& \qquad =\;\eta(0) \left( \log (x) + \gamma  - \log 2 \right) \:+\: {\mathcal{O}}(x)\qquad\qquad\qquad
\qquad \label{rel1} \\
\lefteqn{ \int_{-\infty}^x \eta(a)\: \frac{1}{a}\, \frac{d}{da} \left(a^2 \, \frac{d}{da}\, (\Theta(a)\,\log a)
\right)
\;=\; \eta(0) \left( \log (x) + 1 \right) \:+\: {\mathcal{O}}(x)\:. }
\end{eqnarray}
Thus, choosing~$c$ according to~(\ref{cdef}), we obtain
\[ \lim_{L \rightarrow \infty} \left\{ \frac{1}{a} \left( e^{-\frac{\varepsilon a}{2}}
-e^{-\frac{La}{2}} \right) \Theta(a) - \delta(a) (c-1+\log L) \right\}
\;=\; -\frac{1}{4}\: \hat{W} \! \left( \log(a)\, \Theta(a) \right) e^{-\frac{\varepsilon a}{2}}\:, \]
where~$\hat{W}$ is the differential operator~(\ref{woms}), and the derivatives and the
convergence are meant in~${\mathcal{S}}'(\R)$.
Using at the origin a scaling argument similar to~(\ref{origin}), we can thus take the
limit~$L \rightarrow \infty$ on the left of~(\ref{inid}) to obtain~(\ref{611}).

In order to derive~(\ref{612}), we apply the operator~$-i \Pdd_\xi$ to~(\ref{inid}).
Using~(\ref{rel1}) together with the identity
\[ \int_{-\infty}^x \eta(a)\: \frac{d}{da}\, (\Theta(a)\,\log a)
\;=\; \eta(0) \log (x) \:+\: o(x) , \]
we conclude that
\[ \lim_{L \rightarrow \infty} \left\{ \frac{k\slsh}{a} \left( e^{-\frac{\varepsilon a}{2}}
-e^{-\frac{La}{2}} \right) \Theta(a) - \delta(a) (c+\log L) \right\} \epsilon(k^0)
\;=\; \frac{1}{2}\; \Pdd_k \! \left( \log(a)\, \Theta(a)\, \epsilon(k^0)
\right) e^{-\frac{\varepsilon a}{2}}\:, \]
giving~(\ref{612}).
\QED
The method which we just used to take the limit~$L \rightarrow \infty$ can
be also be applied to take the limit~$\varepsilon \searrow 0$ in~(\ref{611}) and~(\ref{612}),
giving the following result.
\begin{Corollary} \label{lemmaA2}
The following equations hold in the sense of distributions:
\begin{eqnarray}
\int \frac{d^4 k}{(2 \pi)^4}\:\Box_k\! \left( \log(k^2)\, \Theta(k^2)\, \epsilon(k^0)
\right) e^{ik \xi}  &=& -\frac{i}{4 \pi^2}\: \Box_\xi\! \left( \log(\xi^2)\, \Theta(\xi^2)\,
\epsilon(\xi^0) \right) \label{A21} \\
\int \frac{d^4 k}{(2 \pi)^4}\:\Pdd_k\! \left( \log(k^2)\, \Theta(k^2)\, \epsilon(k^0)
\right) e^{ik \xi}  &=& -\frac{1}{8 \pi^2}\: \Pdd_\xi \Box_\xi\! \left((\log(\xi^2)-1)\,
\Theta(\xi^2)\, \epsilon(\xi^0) \right) \qquad \label{A22}
\end{eqnarray}
\end{Corollary}

Let us explain why the previous Proposition is so useful for regularizing~$\tM$
near the light cone. Our goal is to introduce a regularization~$\tM^\varepsilon$ which
is explicit both in position and momentum space and makes our action finite, i.e.\
\beq \label{af}
\int_0^\infty (\tM^\varepsilon)^2\, a\, da \;<\; \infty
\eeq
(see also~(\ref{A0reg}) and the discussion thereafter). Carrying out the derivatives,
one sees that the distribution~(\ref{A22}) has a pole of the desired form~(\ref{leading}),
\[ (\ref{A22}) \;=\; \frac{2i}{\pi^2}\: \frac{\xi\slsh}{\xi^4}\: \Theta(\xi^2)\, \epsilon(\xi^0)
\qquad {\mbox{if $\xi \not \in L$}}. \]
Thus~(\ref{A22}) extends this pole in the distributional sense across the light cone and gives
an explicit formula for the corresponding Fourier transform. But this does clearly not
solve the problem of making the integral~(\ref{af}) finite. The integrand in~(\ref{612})
differs from that in~(\ref{A22}) by the factor~$e^{-\frac{\varepsilon k^2}{2}}$.
The decay of this factor for large~$k^2$ can be thought of as a ``smooth momentum
cutoff,'' and thus we refer to it as the
\beq \label{rm}
{\mbox{regularizing factor~$e^{-\frac{\varepsilon k^2}{2}}$ in momentum space.}}
\eeq
The interesting point is the effect of this regularization on the pole in position space
as described by the right side of~(\ref{612}). The term~$\delta(\xi^2)$ leads to a
distributional contribution supported on the light cone. It is not obvious how to make sense
of it in an integral of the form~(\ref{af}). But we can simply drop this contribution by
considering instead of~(\ref{af}) an integral of the form
\beq \label{disregard}
\int_{0_+}^\infty (\tM^\varepsilon)^2\, a\, da \;:=\;
\lim_{\delta \searrow 0} \int_\delta^\infty (\tM^\varepsilon)^2\, a\, da
\eeq
(we remark that an equivalent, but maybe a bit cleaner method for getting rid of
the distribution supported at~$a=0$ is to subtract this distribution from~(\ref{af})
and to use that, according to~(\ref{A13}), this changes the Fourier transform
only by an irrelevant contribution supported on the mass cone~$\{k \:|\: k^2=0\}$).
Disregarding the $\delta$-distribution, the effect of the regularization can be described by the
\beq \label{rp}
{\mbox{replacement}} \;\; \frac{1}{z^2} \;\longrightarrow\; -\frac{d}{dz}
\left[ \frac{1}{z} \left(1-e^{-\frac{z}{2 \varepsilon}} \right) \right] \;\; {\mbox{in position space}}.
\eeq
Using the formula
\[ \frac{1}{z} \left(1-e^{-\frac{z}{2 \varepsilon}} \right) \;=\;
\int_0^{\frac{1}{2 \varepsilon}} e^{-\lambda z} \, d\lambda\:, \]
one sees that the square bracket in~(\ref{rp}) is a smooth function in the variable~$z$.
Also, due to the decay
of the factor~$e^{-\frac{z}{2 \varepsilon}}$, the regularization is indeed localized in a
strip $z \sim \varepsilon$ around the light cone. With~(\ref{rm}) and~(\ref{rp}) we have
a simple regularization method, which is convenient both in momentum and position space.

We are now ready to complete the constructions outlined in Section~\ref{sec4}.
According to the left of~(\ref{A22}), the pole~$\sim \xi\slsh/\xi^4$ corresponds in momentum
space to a term of the form~$k\slsh/k^2$.  Thus the relevant term for the regularization
of~$\hM$ can be read off from the expansion~(\ref{hMexinf}).
\begin{Def} \label{defhMe} For any~$\varepsilon>0$ and~$k \in {\mathcal{C}}$ with~$k^2>0$,
we introduce the function~$\hN^\varepsilon$ by
\[ \hN^\varepsilon(k) \;=\; \frac{1}{64 \pi^3} \, \frac{k\slsh}{k^4}\, \epsilon(k^0)
\sum_{\alpha, \beta=1}^g \rho_\alpha \,\rho_\beta\: K(k^2, m_\alpha, m_\beta)
\:+\: 2\pi^2\, \m_3 \:\frac{k\slsh}{k^2}
\: \left( e^{-\frac{\varepsilon k^2}{2}} - 1 \right) \epsilon(k^0) \:, \]
where~$K$ is the function~(\ref{Kdef}).
For any~$\varepsilon>0$, we define the distribution~$\hM^\varepsilon$ by
\beq \label{hMepsdef}
\hM^\varepsilon(k) \;=\; \hM(k) + 2\pi^2\, \m_3 \:\frac{k\slsh}{k^2}
\: \left( e^{-\frac{\varepsilon k^2}{2}} - 1 \right) \Theta(k^2)\, \epsilon(k^0)
\eeq
with~$\hM$ according to~(\ref{e1}).
The Fourier transform of~$\hM^\varepsilon$ is denoted by~$\M^\varepsilon$.
\end{Def}

\vspace*{.5em}
\noindent
{\em{Proof of Lemma~\ref{lemmalater}. }}
From the expansion~(\ref{hMex}) we know that for any~$\varepsilon>0$,
the pole of~$\hM^\varepsilon(k)$
on the mass cone is integrable in the~$L^2(\hat{M}, a\, da)$-norm.
Furthermore, from~(\ref{hMexinf}) and Definition~\ref{defhMe} we see
that~$\M^\varepsilon(k)$ for large~$k^2$ has the form
\beq \label{largeasy}
\hM^\varepsilon(k) \;=\;
2\pi^2 \:k\slsh\:\Theta(k^2)\, \epsilon(k^0)
\left(\frac{\m_3}{k^2}\: e^{-\frac{\varepsilon k^2}{2}} + \frac{4 \m_5}{k^4} \right) 
\qquad {\mbox{if~$k^2 > \max_{\beta \in \{1,\ldots, g\}} m_\beta^2$}}\:.
\eeq
We conclude that~$\hM^\varepsilon \in L^2(\hat{M}, a\, da)$, and thus Corollary~\ref{corplan}
applies,
\[ \int_{0_+} \Tr ( (\M^{\varepsilon})^2) \, z\, dz \;=\; \frac{1}{(2 \pi)^4}
\int_{0_+} \Tr ( (\hM^{\varepsilon})^2) \, a\, da \:, \]
were we disregard the distributional contributions at~$z=0$ and~$a=0$ as explained
before~(\ref{disregard}). From~(\ref{largeasy}) we can also determine
the counter term needed in~(\ref{regact}). Namely, substituting~(\ref{largeasy})
into~(\ref{regact}) and computing the integral, we find that choosing
\[ F_{\varepsilon}(\m_3, \m_5) \;=\; G(\m_3, \m_5)
-\frac{4 \pi^4\, \m_3^2}{\varepsilon} \:-\: 32 \pi^4\, \m_3 \m_5  \log\varepsilon \:, \]
the limit~$\varepsilon \searrow 0$ exist. The limit clearly coincides with
that in~(\ref{Smdef}) for a suitable choice of the function~$F-G$.

The identity~(\ref{hMdistr}) follows immediately from Definition~\ref{defhMe},
and thus it remains to justify the limits in~(\ref{dS1}) and~(\ref{dS2}).
Since~$\hF$ vanishes for large~$k^2$ according to~(\ref{Fvanish}), the integral
clearly converges for large~$a$. For the behavior near~$a=0$, we use the
expansion~(\ref{convex}) to rewrite the integral in~(\ref{dS1}) as
\beq \label{distr}
\int_0^\infty \left[ \frac{k\slsh}{a}\: \hM^\varepsilon \right] \Big(c + {\mathcal{O}}(a)
\Big)\, da\:,
\eeq
where~$c$ stands for the leading coefficient in~(\ref{convex}). According to
Definition~\ref{defhMe} and~(\ref{e1}), the square bracket converges in~${\mathcal{S}}'(\R)$,
and therefore~(\ref{distr}) makes sense in the limit~$\varepsilon \searrow 0$ if we interpret
the round brackets as the test function.

To analyze~(\ref{dS2}), we split up the integral into an integral over the interval~$a \in (0,1]$
plus an integral over~$[1, \infty)$. Using the explicit form of~$\hN_\varepsilon$, the integral
over~$(0,1]$ can again be written in the form~(\ref{distr}), which establishes convergence
as~$\varepsilon \searrow 0$. The integral over~$[1,\infty)$, on the other hand, can be analyzed exactly as that in~(\ref{regact}), because~$\hN^\varepsilon$ and~$\hM^\varepsilon$
coincide for large~$a$ according to~(\ref{hMdistr}, \ref{Fvanish}).
\QED

\section{Additional Free Parameters}
\setcounter{equation}{0}  \label{sec7}
In view of the application to state stability, the following consideration gives rise to
additional free parameters. As is shown in~\cite[Appendix~A]{F4}, it is possible to regularize
the fermionic projector in such a way that the property of a distribution~$\M P$-product is
no longer satisfied, but where the continuum limit yields an additional
distributional contributions supported at~$\xi=0$. More precisely, (\ref{tMP}) is replaced by
\[ Q(\xi) \;=\; \frac{1}{2}\: \tM(\xi)\: P(\xi) \:+\: c_2\, \delta^4(\xi) - c_3\: i \Pdd \,\delta^4(\xi)
- c_4\: \Box \delta^(\xi) \]
with real parameters~$c_2$, $c_3$ and~$c_4$, which can be chosen arbitrarily.
Let us consider how these additional parameters can be built into the variational principle
of Definition~\ref{def21}. Taking the Fourier transform, we see that the additional
terms change the functions~$a$ and~$b$ in the representation~(\ref{62f}) by
\beq \label{abchange}
a(k^2) \;\rightarrow\; a(k^2) + c_3\:|k| \:,\qquad b(k^2) \;\rightarrow\; b(k^2)
+ c_2 + c_1\, k^2\:.
\eeq
The parameter~$c_2$ changes~$Q$ only by an irrelevant constant, and thus we
may disregard this parameter. Then, going through the calculation in the
proof of Theorem~\ref{thm1}, one immediately verifies that the terms in~(\ref{abchange})
can be arranged by modifying the action as follows.
\begin{Def} \label{defextact} We define the
{\bf{extended action}} ${\mathcal{S}}_{\mbox{\scriptsize{ext}}}$ by
\[ {\mathcal{S}}_{\mbox{\scriptsize{ext}}} \;=\; {\mathcal{S}}
\:+\: c_3 \sum_{\beta=1}^g \rho_\beta\, m_\beta^4 \:+\: c_4
\sum_{\beta=1}^g \rho_\beta\, m_\beta^5 \]
with~${\mathcal{S}}$ as in~(\ref{Sdef}). Here~$c_3$ and~$c_4$ are two
free real parameters. The corresponding variational principle with constraint~(\ref{constraint})
is referred to as the {\bf{extended variational principle}}.
\end{Def}

\section{Numerical Construction of Minimizers}
\setcounter{equation}{0}  \label{sec8}
We first rewrite the action in a form which is most convenient for the numerical analysis.
\begin{Prp} \label{prp81} {\bf{(Extended action in momentum space)}}
We introduce the real-valued function
\beq \label{Hdef}
H(a,x,y) \;=\; \frac{1}{a} \: \Big( J(a,x,y) + K(a,x,y) \Big)
\eeq
(with the functions~$J$ and~$K$ as given by~(\ref{Jdef}, \ref{Kdef}))
and choose a parameter
\[ a_{\max} \;>\; \max_{\beta \in \{1,\ldots, g\}} m_\beta^2 \:. \]
Then the extended action (see Definitions~\ref{defextact} and~\ref{def21}) can be written as
\begin{eqnarray}
{\mathcal{S}}_{\mbox{\scriptsize{ext}}} &=& \frac{1}{2^{16} \pi^{10}}
 \sum_{\alpha, \beta, \gamma, \delta} \rho_\alpha\, \rho_\beta\, \rho_\gamma\, \rho_\delta
\int_0^{a_{\max}}H(a, m_\alpha, m_\beta)\: H(a, m_\gamma, m_\delta)\: da 
\label{Sim1} \\
&&+\: F(a_{\max}, \m_3, \m_5)
\:+\: c_3 \sum_{\beta=1}^g \rho_\beta\, m_\beta^4 \:+\: c_4
\sum_{\beta=1}^g \rho_\beta\, m_\beta^5\:. \label{Sim2}
\end{eqnarray}
\end{Prp}
{\Proof} It clearly suffices to consider~${\mathcal{S}}$, because the extended
action is obtained simply by adding the extra terms in Definition~\ref{defextact}.
From~(\ref{hMepsdef}) and~(\ref{hMexinf}) it is clear that~$\hM^\varepsilon(k)
\stackrel{\varepsilon \rightarrow 0}{\rightarrow} \hM(k)$ uniformly on the set $\{k \:|\:
0 < k^2 < a_{\max} \}$, whereas for large~$k^2$, $\hM^\varepsilon$ is given explicitly by
\[ \hM^\varepsilon(k) \;=\; 2\pi^2 \:k\slsh\:\Theta(k^2)\, \epsilon(k^0)
\left(\frac{\m_3}{k^2}\: e^{-\frac{\varepsilon k^2}{2}} + \frac{4 \m_5}{k^4} \right) \qquad
{\mbox{if~$k^2 > a_{\max}$}}\:. \]
Hence, using the result of Theorem~\ref{thm58} in the formula for the
regularized action in momentum space~(\ref{regact}), we obtain
\begin{eqnarray*}
{\mathcal{S}} &=& \frac{1}{2^{16} \pi^{10}}
 \sum_{\alpha, \beta, \gamma, \delta} \rho_\alpha\, \rho_\beta\, \rho_\gamma\, \rho_\delta
\int_0^{a_{\max}}H(a, m_\alpha, m_\beta)\: H(a, m_\gamma, m_\delta)\: da \\
&&+ \frac{1}{(2 \pi)^4} \lim_{\varepsilon \searrow 0} \left( 4 \pi^4 \int_{a_{\max}}^\infty
\left(\m_3 \,e^{-\frac{\varepsilon k^2}{2}} + 4 \m_5\, a^{-1} \right)^2
da + F_\varepsilon(\m_3, \m_5) \right) .
\end{eqnarray*}
The result follows because the last line depends only on~$a_{\max}$,
$\m_3$ and~$\m_5$.
\QED
Note that the argument before~(\ref{hMex}) shows that the function~$H$, (\ref{Hdef}), is
bounded near~$a=0$. Hence in~(\ref{Sim1}) we only need to integrate
a bounded continuous function over a compact interval. This makes it possible to
compute the action with standard numerical methods. Using standard tools of nonlinear
optimization, one can construct numerical minimizers.

Alternatively, one can construct critical points by searching for numerical solutions of the
conditions~(a) and~(b) in Theorem~\ref{thm1}. It is helpful that the resulting equations
are polynomials in the parameters~$\rho_\beta$, making it possible to use specific algorithms
for computing the roots of systems of polynomials.
For a detailed description of the numerics we refer to the PhD thesis~\cite{H}; here
we proceed by discussing a few of the numerical solutions.

\begin{figure}[tb]
\begin{center}
{\includegraphics{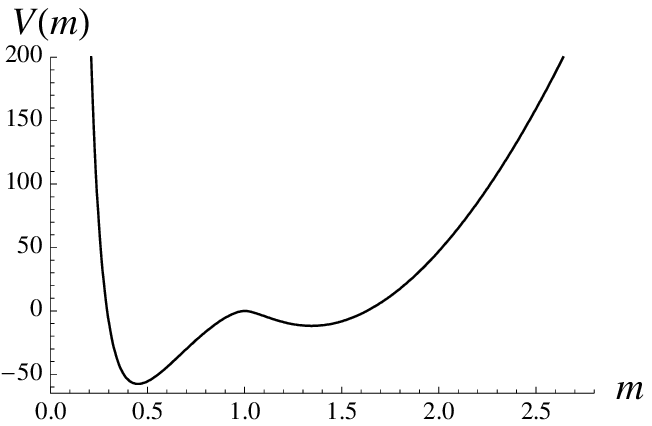}}
{\includegraphics{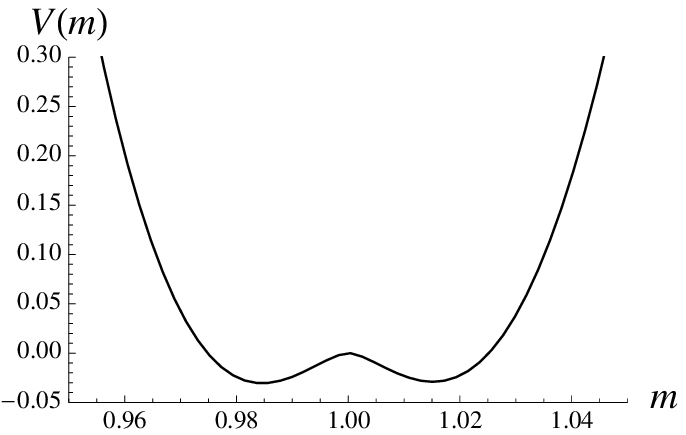}}
\caption{The variation density in the case of one generation of mass~$m_1=1$
and the parameters~$c_0=c_3=c_4=0$ and~$c_1=2 \cdot 10^6$ (left),
$c_1=10^7$ (right).}
\label{figone}
\end{center}
\end{figure}

In the case~$g=1$ of one generation, we constructed all minimizers of the
unextended action (i.e.\ $c_3=c=4=0$). By scaling we can assume that~$m_1=1$.
We found a one-parameter family of minimizers parametrized by~$c_1$, whereas~$c_0$
must vanish. In Figure~\ref{figone} the variation density is shown in
typical examples.
In agreement with Theorem~\ref{thm1} and Theorem~\ref{thm2}, we see that $V'(1)=0$.
However, the solutions are not state stable, because the condition~(iii) in Definition~\ref{def611}
(or equivalently the condition~(iii') on page~\pageref{pageiii}) is violated.
More specifically, the variation density has near~$m=1$ the expansion
\[ V(m) \;=\; (m-1)^2 \left( \alpha c_1 + \beta \log |m-1| \right) \:+\: o((m-1)^2) \]
with constants~$\alpha, \beta>0$. Therefore, it is impossible to arrange that~$m=1$
is a local minimum of~$V$. Since the extra contributions of the extended action are polynomials
in~$m$, this statement remains true for the extended action. We conclude that
there is no state-stable Dirac sea configuration with one generation. In words, our action
principle prefers to ``split up'' the Dirac sea into two seas. However, by choosing~$c_1$
sufficiently large we can arrange that the minima of~$V$ are arbitrarily close to~$m=1$
(see Figure~\ref{figone} right).

\begin{figure}[tb]
\begin{center}
{\includegraphics{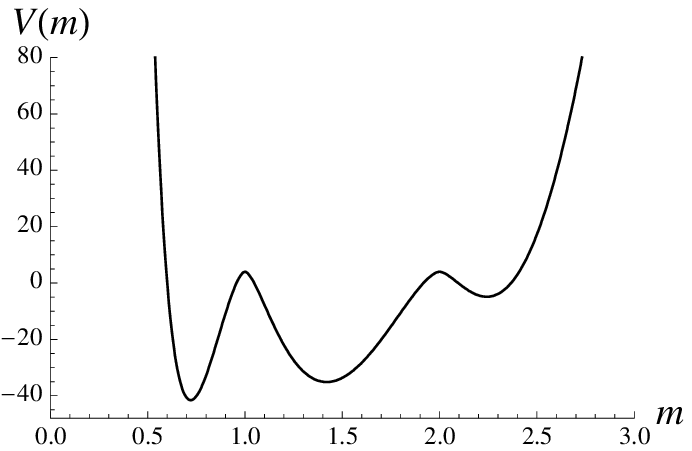}}
{\includegraphics{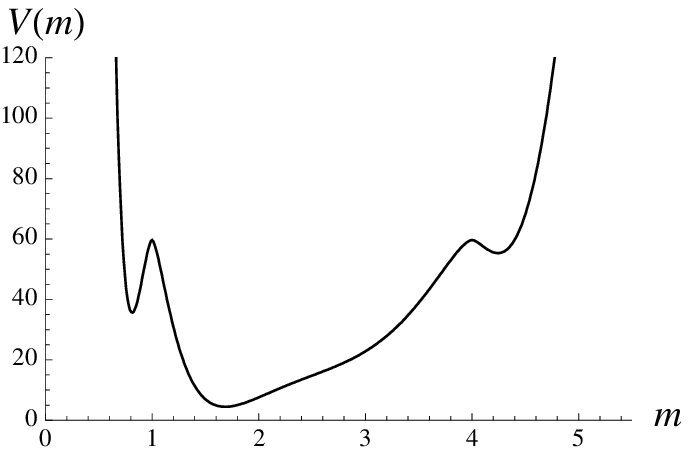}}
{\includegraphics{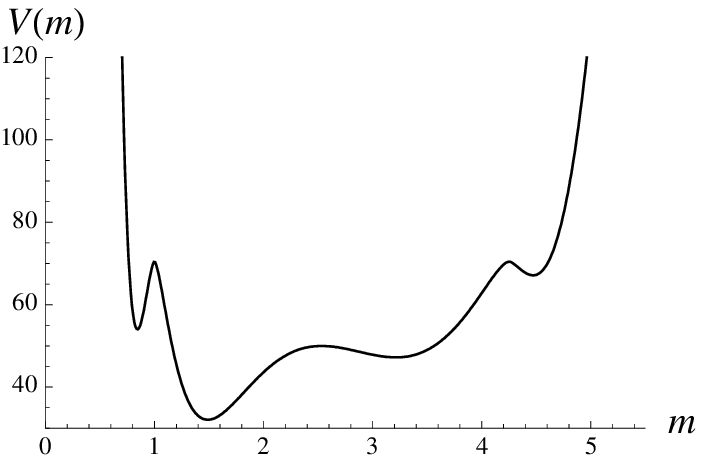}}
{\includegraphics{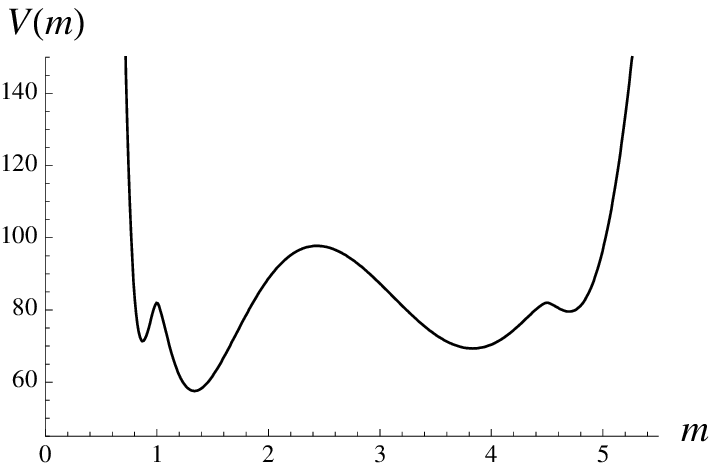}}
{\includegraphics{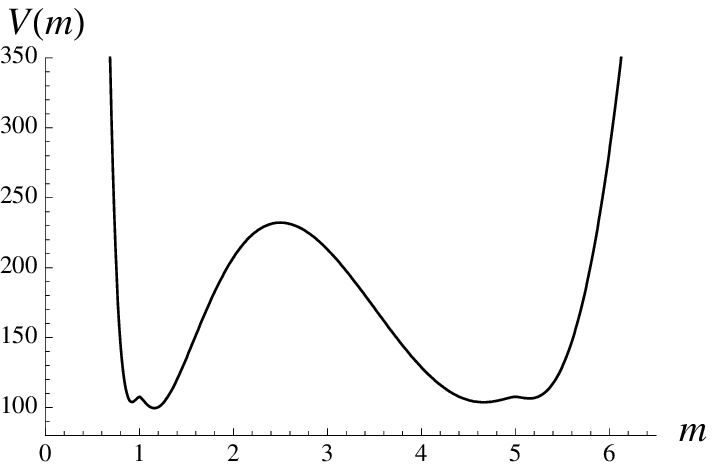}}
{\includegraphics{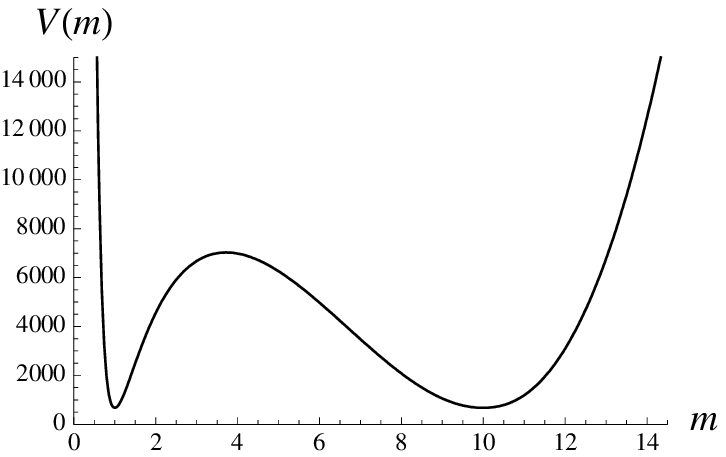}}
\caption{Minimizers of the unextended action in the case of two generations of masses~$m_1=1$
and~$m_2=2, 4, 4.25, 4.5, 5$ and~$10$.}
\label{figtwo}
\end{center}
\end{figure}

The case~$g=2$ of two generations gives some insight into the nature of our variational
principle. By scaling we can always assume that~$m_1=1<m_2$ and~$\rho_1=1$. Thus
for the unextended action we have the four free parameters~$m_2$, $\rho_2$, $c_0$ and~$c_1$
to satisfy the three conditions~$V'(m_1)=0=V'(m_2)$ and~$V(m_1)=V(m_2)$.
Therefore, one expects for given~$m_2$ a discrete number of minima. Indeed, we found
for every~$m_2$ exactly one minimizer. In Figure~\ref{figtwo} the variation density
of the minimizers is shown for different values of~$m_2$.
If~$m_2$ is close to~$m_1$, the result can be understood in analogy to the
minimizers for one generation: again the points~$m=m_j$ are local maxima of the
variation density. Thus the solution is not state stable, and our action principle has the
tendency to split up the two Dirac seas into even more seas.
However, if~$m_2$ is increased, a ``potential wall'' emerges between~$m_1$ and~$m_2$,
such that the number of local minima of~$V$ increases from three to four (see the cases~$m_2=4.25$ and~$m_2=4.5$). The points~$m=m_j$ are still local maxima,
but the neighboring local minima are very close (see the case~$m_2=5$).
If~$m_2$ is further increased, the local minima approach the points~$m=m_j$
exponentially fast, and thus the resulting variation density looks very much
like a state stable configuration (see the case~$m_2=10$).
Keeping in mind that the physical masses of the three
generations of leptons scale with factors of the order~$100$, and that a relative uncertainty in
the rest masses of~$10^{-100}$ can certainly not be measured in experiments, in such
cases it is fair to identify the points~$m_j$ with the neighboring minima, so that we may
regard the~$m_j$ as absolute minima of~$V$.

\begin{figure}[tb]
\begin{center}
{\includegraphics{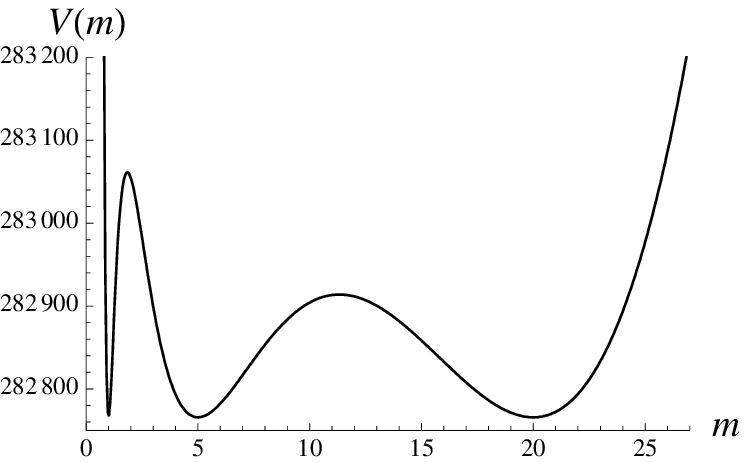}}
{\includegraphics{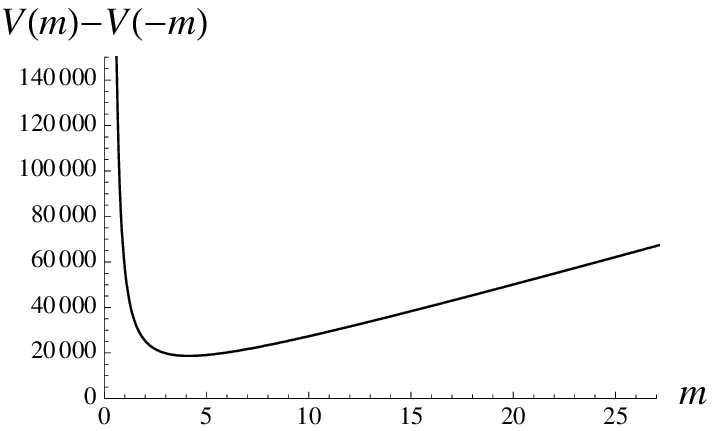}}
\caption{A state stable Dirac sea structure with three generations}
\label{figthree}
\end{center}
\end{figure}

In the case~$g=3$ of three generations, we can assume by scaling that~$1=m_1 < m_2 < m_3$.
We found different classes of minimizers. In view of the physical masses of the leptons,
we were most interested in constructing solutions where the masses are far apart, i.e.\
$m_1 \ll m_2 \ll m_3$. Our attempts to construct such solutions were successful only
for the extended action. Figure~\ref{figthree} gives an example of a state stable
Dirac sea configuration with three generations.
The corresponding values of the parameters are~$m_2=5$, $m_3=20$, $\rho_1=1$,
$\rho_2=10^{-4}$, $\rho_3=9.696 \cdot 10^{-6}$, $c_0=-6.692 \cdot 10^8$,
$c_1=-2.516 \cdot 10^9$, $c_3= 9658.25$, $c_4=8416.56$.

As already mentioned in the introduction, the systematic study of the minimizers
goes beyond the scope of this paper. Clarifying the detailed structure of the minimizers 
and critical points of our action is an important project for the future.
Talking of future developments, we finally point out
that information on the constants~$c_0$, $c_1$, $c_2$ and~$c_4$ could be obtained by
studying regularized or discretized Dirac sea systems (for example using numerical
methods~\cite{FP} for large systems).
Moreover, the weight factors~$\rho_\beta$ have an influence on the
continuum limit (see~\cite[Appendix~A]{F4}), and thus the detailed analysis
of the continuum limit for interacting systems should give constraints for the weight factors.
In view of these future investigations, which should reduce our number of free parameters,
the introduced action principle seems to be a promising step towards a physical
theory which makes predictions for the ratios of the masses of elementary particles.

\noindent
NWF I -- Mathematik,
Universit{\"a}t Regensburg, 93040 Regensburg, Germany, \\
{\tt{Felix.Finster@mathematik.uni-regensburg.de}}, \\
{\tt{Stefan.Hoch@mathematik.uni-regensburg.de}}

\end{document}